\documentclass[a4paper,11pt]{report}
\usepackage{graphicx,amssymb,amstext,amsmath,url}
\usepackage{palatino,fancyhdr,longtable,lscape,natbib,appendix,floatflt}
\newcommand\micron{\,$\mu$m}

\newcommand\arcsec{$^{\prime\prime}$}
\newcommand{\farcs}{\mbox{\ensuremath{.\!\!^{\prime\prime}}}}
\newcommand\aj{\rmfamily{AJ}}%
\newcommand\apjs{\rmfamily{ApJS}}%
\newcommand\memras{\rmfamily{Memoirs of the Royal Astronomical Society}}%
\newcommand\mnras{\rmfamily{MNRAS}}%
\newcommand\pasj{\rmfamily{PASJ}}%
\newcommand\pasp{\rmfamily{PASP}}%

\begin{document}
\title{{\Huge{SAGE-Spectroscopy:}}\\ The life-cycle of dust and gas in
  the\\ Large Magellanic Cloud\\ {\Large{(PI: F. Kemper, PID: 40159)}}
  ~\\~\\ Data delivery document v3.0} \date{Spring 2011}
\author{Paul~M.~Woods, \and G.C.~Sloan, \and Karl~D.~Gordon, \and
  B.~Shiao, \and F.~Kemper, \and R.~Indebetouw \and \& \and the
  SAGE-Spec~team} \maketitle
\tableofcontents
\newpage
\chapter[Overview]{SAGE-Spec overview}

\section[Introduction to SAGE-Spec]{Introduction to the goals of the SAGE-Spec program}

The SAGE-Spectroscopy legacy proposal (SAGE-Spec; PI: F. Kemper, PID:
40159; Kemper et al. 2010) is the spectroscopic follow-up to the
successful SAGE-LMC legacy program (Meixner et al. 2006; PI:
M. Meixner, PID: 20203) that mapped the Large Magellanic Cloud (LMC)
with all bands of the IRAC and MIPS instruments on board the Spitzer
Space Telescope. The acronym SAGE stands for Surveying the Agents of
Galaxy Evolution, and thus the project aims to make an inventory of
the budget of gas and dust involved in the life-cycle of matter in the
Magellanic Clouds.

The SAGE-Spec legacy program has used the IRS and the SED mode on the
MIPS instrument to obtain spectroscopic information of a
representative sample of point sources and extended regions in the
LMC, with two goals in mind:\\ The first goal is to study the
composition and properties of gas and dust in environments relevant to
the life-cycle of matter, from evolved stars to the interstellar
medium and ultimately to star-forming regions. Combining this
information with statistics obtained from the SAGE-LMC project will
allow us to get a complete picture of the cycle of gas and dust in
terms of mass \emph{and} in terms of composition and physical
properties.\\ Secondly, sources in the SAGE-LMC point source catalogue
will be classified using a classification scheme tested against the
SAGE-Spec sample. In addition to the observations made within the
context of the SAGE-Spec proposal, we have extended the sample to
contain \emph{all} IRS and MIPS-SED observations within the original
SAGE-LMC footprint (\textquotedblleft the archival
sample\textquotedblright).

In total, 224.6 hours of observations were made and 196 positions were
targeted with the IRS instrument, resulting in 197 point sources being
observed; 48 point sources were observed with MIPS in SED mode and 20
extended regions were also targeted with MIPS: 10 H{\sc{ii}} regions
and 10 diffuse regions. In addition, the reduction and delivery of all
the IRS and MIPS-SED spectroscopic data within the SAGE-LMC footprint
adds a further $\sim$900 IRS (mostly point-source) targets and several
other extended regions. See \citet{kem10} for an overview of the
project.

Of the 196 point sources observed with IRS, all were observed using
the Short-Low module (SL; $\lambda\approx$~5--14\,$\mu$m), and 128
were also observed using Long-Low (LL;
$\lambda\approx$~14--37\,$\mu$m). Upon examination, the LL observation
of one source, SSID16 (GV~60), picked up another target rather than
the intended target. Thus there are 197 distinct point sources in the
proposal sample at this point. Spectra of a number of serendipitous
objects were obtained, and these will be delivered in a future data
release. 48 of the brightest points sources were also targeted with
MIPS in SED mode and a number of serendipitous sources were also
found, bringing the total of MIPS-SED point sources up to 63 . For our
source selection strategy, please see the overview publication for the
project, Kemper et al. (2010).

\section[Current data release]{Description of the observations in the current data release}

The final SAGE-Spec delivery contains all the reduced IRS staring data
from the Spitzer archive within the SAGE footprint
(Table~\ref{archirs}), totalling 883 spectra. This includes
calibration data. This marks a complete delivery of IRS staring data
from in-orbit checkout through to the end of Cycle 5. We also include
higher-level products: integrated spectra for all IRS and MIPS-SED
extended region data, and the object classification for the original
197 SAGE-Spec IRS point sources, as discussed in \citet{woo11}.

This delivery also includes several updates. Previously-delivered IRS
staring data from the SAGE-Spec program have been updated to the
S18.18 standard. We also deliver revised reduced data of IRS and
MIPS-SED extended regions, accounting for the recently-identified
\textquotedblleft dark settle\textquotedblright\ problem (see
\S\ref{IRS-Map-QC}).

As a bonus, we also include various reduced IRS staring spectra from
the programs found in Table~\ref{archirs} which fall outside of the
SAGE footprint. These have SSIDs of 5000+, and do not form part of the
official SAGE-Spec delivery.

\section[Previous data releases]{Description of the observations in previous data releases}

\subsection{Delivery 2}

The second delivery of data from the SAGE-Spec Spitzer Legacy Program
improved upon the data delivered in delivery \#1 and added both IRS
and MIPS-SED data on point sources and extended regions observed in
the LMC with Spitzer. We delivered reduced data from the Spitzer
archive that falls within the SAGE footprint on the LMC. This delivery
included IRS staring mode data from Spitzer cycles 1--3 and the
Guaranteed Time Observations (GTOs). It comprised 403 spectra of 352
objects or lines of sight. Details of program data included can be
found in Table~\ref{archirs}. We matched these spectral data with
near- and mid-IR photometry (see below). In addition, we updated some
of the matching broadband photometry previously delivered with the 197
SAGE-Spec proposal point sources.

We also released IRS mapping data of 23 diffuse and H{\sc ii} regions
in the form of order cubes, 3 of which were taken from the Spitzer
archive. All but one (SSDR12) of these regions are also covered by
MIPS-SED observations, for which we also delivered reduced data. Our
previous reductions of MIPS-SED data of 48 point sources were revised,
and several serendipitous sources were added to the sample so that the
number now totals 61 (with 63 spectra; Table~\ref{mipssed}).

\subsection{Delivery 1}

In the first delivery of SAGE-Spec data in August 2009 we delivered
IRS staring mode spectra of 197 point sources and MIPS-SED spectra of
48 point sources selected from the SAGE catalogue. The IRS staring
mode spectra are complemented by matching photometry from SAGE
\citep[IRAC and MIPS;][]{mei06}, MCPS \citep{zar04}, 2MASS
\citep{skr06} and IRSF \citep{kat07}. We also delivered Postscript
plots of the 197 the IRS spectra with associated photometry.

\section[Future data releases; timeline]{Description of the observations in future data releases and delivery timeline}

Further data releases from the SAGE-Spec project will include:
\begin{itemize}
\item Object classifications for the entire Spitzer IRS archive
located within the SAGE footprint (GTO, cycles 1--5, calibration
data).
\item Spectral classifications for the entire Spitzer IRS archive
located within the SAGE footprint (GTO, cycles 1--5, calibration
data).
\item Statistical object
classification of the SAGE point source catalogue.
\item Feature maps of
extended regions covered by IRS and MIPS-SED observations.
\end{itemize}
Minor update deliveries may be made for revised data reductions. Due
to the closure of the Spitzer Science Centre, these deliveries will be
made to IRSA directly, and/or presented in journal articles.

\begin{longtable}{|l|c|c|c|}
\caption{Archival IRS staring data in the present delivery} \label{archirs} \\

\hline \multicolumn{1}{|c|}{\textbf{Cycle}} & \multicolumn{1}{c|}{\textbf{Program ID}} & \multicolumn{1}{c|}{\textbf{Program PI}} & \multicolumn{1}{|c|}{\textbf{notes}} \\ \hline 
\endfirsthead

\multicolumn{4}{c}%
{{\bfseries \tablename\ \thetable{} -- continued from previous page}} \\
\hline \multicolumn{1}{|c|}{\textbf{Cycle}} & \multicolumn{1}{c|}{\textbf{Program ID}} & \multicolumn{1}{c|}{\textbf{Program PI}} & \multicolumn{1}{|c|}{\textbf{notes}} \\ \hline 
\endhead

\hline \multicolumn{4}{r}{{continued on next page\ldots}} \\ \hline
\endfoot
\hline \hline
\endlastfoot
Calsfx& M -- 28  & Various & 14 hires, 43 lores spectra\\
1 (GTO) & 18    & Uchida        & 10 hires spectra\\
1 (GTO) & 63    & Brandl        & 19 hires, 17 lores spectra \\
1 (GTO) & 103   & Bernard-Salas & 18 hires, 18 lores spectra\\
1 (GTO) & 124, 129 & Gehrz      & 6 slhi, 2 lores spectra\\
1 (GTO) & 200   & Sloan         & 29 lores spectra\\
1 (GTO) & 249   & Indebetouw    & 6 lores spectra\\
1 (GTO) & 263   & Woodward      & 1 lores spectrum\\
1 (GTO) & 464, 472   & Cohen    & 14 lores spectra\\
1 (GTO) & 485   & Ardila        & 6 lores spectra\\
1   & 1094  & Kemper        & 9 hires, 2 slhi, 9 lores spectra \\
1   & 2333  & Woodward      & 4 slhi spectra\\
1   & 3426  & Kastner       & 60 lores spectra\\
1   & 3470  & Bouwman       & 1 hires spectrum\\
1   & 3483  & Rho           & 1 lores spectrum\\
1   & 3505  & Wood          & 31 lores spectra\\
1   & 3578  & Misselt       & 13 lores spectra\\
1   & 3583  & Onaka         & 12 lores spectra\\
1   & 3591  & Kemper        & 53 lores spectra\\
2   & 20080 & Polomski      & 1 lores spectrum\\
2   & 20443 & Stanghellini  & 20 lores spectra\\
2   & 20752 & Reynolds      & 1 hires spectrum\\
3   & 30067 & Dwek          & 4 lores spectra\\
3   & 30077 & Evans         & 1 lores spectrum\\
3   & 30180 & IRAC          & 1 lores spectrum\\
3   & 30332 & Sloan         & 1 lores spectrum\\
3   & 30345 & Sloan         & 4 hires spectra\\
3   & 30372 & Tappe         & 7 lores spectra\\
3   & 30380 & Clayton       & 1 lores spectrum\\
3   & 30544 & Morris        & 1 lores spectrum\\
3   & 30788 & Sahai         & 24 lores spectra\\
3   & 30869 & Kastner       & 1 slhi, 8 lores spectra\\
4   & 40031 & IRAC          & 2 lores spectra\\
4   & 40149 & Dwek          & 4 lores spectra\\
4   & 40604 & Reynolds      & 2 lores spectra\\
4   & 40650 & Looney        & 8 hires, 331 slhi spectra\\
5   & 50092 & Gielen        & 10 lores spectra\\
5   & 50147 & Sloan         & 6 hires spectra\\
5   & 50167 & Clayton       & 7 lores spectra\\
5   & 50338 & Matsuura      & 11 hires, 1 slhi, 23 lores spectra\\
5   & 50444 & Dwek          & 3 hires, 2 lores spectra\\
\end{longtable}

\begin{longtable}{|l|l|c|c|c|}
\caption{Point sources observed in MIPS-SED mode} \label{mipssed} \\
\hline \multicolumn{1}{|c|}{\textbf{SSID}} & \multicolumn{1}{c|}{\textbf{Source name}} & \multicolumn{1}{c|}{\textbf{Co-ordinates (J2000)}} & \multicolumn{1}{|c|}{\textbf{AOR key}} & \multicolumn{1}{c|}{\textbf{Deliv.\#2 SSID}}\\ \hline 
\endfirsthead

\multicolumn{5}{c}%
{{\bfseries \tablename\ \thetable{} -- continued from previous page}} \\
\hline \multicolumn{1}{|c|}{\textbf{SSID}} & \multicolumn{1}{c|}{\textbf{Source name}} & \multicolumn{1}{c|}{\textbf{Co-ordinates (J2000)}} & \multicolumn{1}{|c|}{\textbf{AOR key}} & \multicolumn{1}{c|}{\textbf{Former SSID}}\\ \hline 
\endhead

\hline \multicolumn{5}{r}{{continued on next page\ldots}} \\ \hline
\endfoot

\hline \hline
\caption{The reader is also referred to \citet{vlo10} for integration times, quality control and other information.}
\endlastfoot

600 & SAGE1C J043727.60-675435.0     & 4h37m27.61s, -67d54m35.0s & 22459648 & 500 \\
601 & SMP-LMC-11                     & 4h51m37.82s, -67d05m17.0s & 22450432 & 501 \\
602 & IRAS04530-6916                 & 4h52m45.70s, -69d11m49.4s & 22453504 & 502 \\
603 & IRAS04537-6922                 & 4h53m30.11s, -69d17m49.3s & 22457088 & 503 \\
604 & LH$\alpha$120-N89              & 4h55m06.53s, -69d17m08.5s & 22460416 & 504 \\
605 & WOH 6064                       & 4h55m10.48s, -68d20m29.9s & 22448640 & 505 \\
606 & IRAS04557-6639                 & 4h55m50.59s, -66d34m34.7s & 22457344 & 506 \\
607 & IRAS04562-6641                 & 4h56m22.59s, -66d36m56.8s & 22457600 & 507 \\
608 & R66                            & 4h56m47.08s, -69d50m24.8s & 22453760 & 508 \\
609 & R71                            & 5h02m07.40s, -71d20m13.0s & 22454016 & 509 \\
610 & IRAS05047-6644                 & 5h04m47.00s, -66d40m30.3s & 22451456 & 510 \\
611 & SMP-LMC-21                     & 5h04m51.97s, -68d39m09.5s & 22450688 & 511 \\
612 & SMP-LMC-28                     & 5h07m57.62s, -68d51m47.3s & 22450944 & 512 \\
613 & SMP-LMC-36                     & 5h10m39.60s, -68d36m04.9s & 22459664 & 513 \\
614 & IRAS05137-6914                 & 5h13m24.66s, -69d10m48.1s & 22452992 & 514 \\ 
615 & MSX-LMC222                     & 5h13m41.99s, -69d35m26.7s & 22451712 & 515 \\
616 & SAGE1C J051347.73-693505.3     & 5h13m46.5s,  -69d35m10s   & 22451712 & --- \\
617 & MSX-LMC349                     & 5h17m26.93s, -68d54m58.7s & 22449920 & 516 \\
618 & IRAS05216-6753                 & 5h21m29.68s, -67d51m06.6s & 22451968 & 517 \\
619 & SAGE1C J052222.95-684101.0     & 5h22m22.96s, -68d41m01.1s & 22457856 & 518 \\
620 & HS270-IR1                      & 5h23m53.93s, -71d34m43.8s & 22458112 & 519 \\
621 & SMP-LMC-62                     & 5h24m55.08s, -71d32m56.1s & 22451200 & 520 \\
622 & LH$\alpha$120-N51-YSO1         & 5h26m01.22s, -67d30m11.9s & 22458368 & 521 \\
623 & LH$\alpha$120-N49              & 5h26m03.10s, -66d05m17.2s & 22454272 & 522 \\
624 & LH$\alpha$120-N49              & 5h26m07s,    -66d05m00s   & 22454272 & --- \\
625 & MSX-LMC577                     & 5h26m30.60s, -67d40m36.7s & 22455296 & 523 \\
626 & LH55                           & 5h26m38.6s,  -67d39m23s   & 22455296 & --- \\ 
627 & IRAS05281-7126                 & 5h27m23.14s, -71d24m26.3s & 22458624 & 524 \\
628 & IRAS05280-6910                 & 5h27m40.06s, -69d08m04.6s & 22454528 & 525 \\
629 & IRAS05291-6700                 & 5h29m07.59s, -66d58m15.1s & 22450176 & 526 \\
630 & IRAS05298-6957                 & 5h29m24.53s, -69d55m15.9s & 22448896 & 527 \\
631 & IRAS05325-6629                 & 5h32m31.95s, -66d27m15.2s & 22452224 & 528 \\
632 & IRAS05328-6827                 & 5h32m38.59s, -68d25m22.4s & 22452480 & 529 \\
633 & RP775                          & 5h32m44.40s, -69d30m05.5s & 22459904 & 530 \\
634 & IRAS05330-6826                 & 5h32m48.3s,  -68d23m59s   & 22452480 & --- \\
635 & SAGE1C J053249.35-693036.8     & 5h32m49.4s,  -69d30m37s   & 22459904 & --- \\
636 & IRAS05329-6708                 & 5h32m51.36s, -67d06m51.8s & 22449152 & 531 \\
637 & MSX-LMC783                     & 5h32m55.44s, -69d20m26.6s & 22455552 & 532 \\
638 & HV2671                         & 5h33m48.92s, -70d13m23.6s & 22454784 & 533 \\
639 & SN 1987A                       & 5h35m23.8s,  -69d17m07s   & 17720320 & --- \\
640 & MSX-LMC741                     & 5h35m25.83s, -71d19m56.6s & 22455808 & 534 \\
641 & SN 1987A                       & 5h35m31.5s,  -69d17m09s   & 22393344 & --- \\
642 & SN 1987A                       & 5h35m38.1s,  -69d16m43s   & 17721600 & --- \\
643 & SN 1987A                       & 5h35m38.2s,  -69d16m43s   & 22394624 & --- \\
644 & R126                           & 5h36m25.85s, -69d22m55.7s & 22455040 & 535 \\
645 & 30Dor-17                       & 5h37m28.09s, -69d08m47.8s & 22452736 & 536 \\
646 & H88b 86?                       & 5h38m40.9s,  -69d25m14s   & 22460672 & --- \\
647 & LH$\alpha$120-N158B            & 5h38m44.53s, -69d24m38.3s & 22460672 & 537 \\
648 & LH$\alpha$120-N159-P2          & 5h39m41.86s, -69d46m11.9s & 22453248 & 538 \\
649 & LH$\alpha$120-N159-K4?         & 5h39m55.0s,  -69d46m16s   & 22459136 & --- \\
650 & LH$\alpha$120-N160-1           & 5h39m59.49s, -69d37m30.3s & 22458880 & 539 \\
651 & LH$\alpha$120-N159S            & 5h40m00.67s, -69d47m13.4s & 22459136 & 540 \\
652 & LH$\alpha$120-N160-3           & 5h40m04.4s,  -69d38m21s   & 22458880 & --- \\
653 & UFO1                           & 5h40m11.83s, -70d10m04.2s & 22456064 & 541 \\
654 & WOH-G457?                      & 5h40m11.95s, -70d09m15.7s & 22449408 & 542 \\
655 & RP85                           & 5h40m33.57s, -70d32m40.3s & 22460160 & 543 \\
656 & SAGE1C J054043.15-701110.3     & 5h40m43.18s, -70d11m10.3s & 22456320 & 544 \\
656 & SAGE1C J054043.15-701110.3     & 5h40m43.18s, -70d11m10.3s & 22456832 & --- \\
657 & MSX-LMC1794                    & 5h40m44.00s, -69d25m54.5s & 22456576 & 545 \\
658 & MSX-LMC956                     & 5h40m49.27s, -70d10m13.6s & 22456832 & 546 \\
658 & MSX-LMC956                     & 5h40m49.27s, -70d10m13.6s & 22456320 & --- \\
659 & BSDL2955                       & 5h45m43.9s,  -67d09m55s   & 22459392 & --- \\
660 & BSDL2959                       & 5h45m44.80s, -67d09m28.2s & 22459392 & 547 \\
\end{longtable}

\section[Database format]{Format of the database at the Spitzer Science Center}

The SAGE-Spec database can be reached via the Legacy page on the
Spitzer Science Center website, or via direct links:\\
\url{http://irsa.ipac.caltech.edu/data/SPITZER/SAGE/}\\
\url{http://ssc.spitzer.caltech.edu/legacy/sagespechistory.html}

The database is made up of a tables of metadata and associated data
products (FITS files, FITS data cubes, plots of spectra and associated
photometry). The format of these tables is slightly different for the
IRS staring data and the MIPS-SED data.

\subsection{IRS staring point source data}

The IRS staring spectra have an associated metadata file, which is in
IPAC tbl format.This table has slightly different formats for the
SAGE-Spec proposal targets (delivered to the SSC in August 2009) and
the Spitzer archival targets (first delivered to the SSC in February
2010).

\noindent\textbf{SAGESpecID}: This is a unique identifier for each
target. SAGE-Spec point sources are labelled with numbers 1--197,
which are obtained by ordering the sources sequentially according to
Right Ascension (R.A.). Archival sources in this delivery are labelled
4000--4817.\\ \textbf{label}: This is a unique identifier for each
spectrum. It is constructed by concatenating R.A. (to 4 decimal
places) and declination (dec; to 4 decimal places), instrument name
(i.e., IRSX or MIPS) and module name with nod or processing
information, for example:\\ 79.7943-69.5629\_IRSX\_SL\_(scaled).\\ For
the archival targets, this label also includes the SAGE-Spec ID and an
integer which increments with every associated spectrum, for
example:\\ 70.1187-69.9202\_IRSX\_1006\_1\_SL\_LL\_(scaled)\\
70.1190-69.9203\_IRSX\_1006\_2\_SH\_LH\_(scaled)\\
70.1192-69.9206\_IRSX\_1006\_3\_LL\_(scaled)\\ For further details,
see \textbf{modules}, below).\\ \textbf{ra}: Right ascension
Field-Of-View (FOV) co-ordinates
(J2000)\footnotemark.\addtocounter{footnote}{-1}\\ \textbf{dec}:
Declination FOV co-ordinates (J2000)\footnotemark.\footnotetext{We
purposefully choose to use FOV co-ordinates (i.e., ra\_fov and
dec\_fov in the FITS headers) over requested co-ordinates (ra\_rqst
and dec\_rqst, also found in FITS headers, and which are those
obtained when one queries via Leopard, for example) since the FOV
co-ordinates most accurately determine the region of space the
telescope is pointing at. In general, FOV co-ordinates differ from
requested co-ordinates by $\approx$0\farcs75. However, in extreme
cases, there can be a difference of up to $\approx$5\farcs7 between
the two co-ordinate systems in the SAGE-Spec IRS sample.}\\
\textbf{SAGEIRACID}: Associated SAGE-LMC IRAC source designation.\\
\textbf{SAGEMIPSID}: Associated SAGE-LMC MIPS 24\,$\mu$m source
designation.\\ \textbf{AORkey}: Astronomical Observation Request (AOR)
number of the Spitzer observation.\\ \textbf{progid}: Program ID in
which data were obtained (for SAGE-Spec this is 40159).\\
\textbf{instrument}: IRSX for Infrared Spectrograph, MIPS for
Multiband Imaging Photometer for Spitzer.\\
\noindent\textbf{modules}: One of:
\begin{description}
\item SL, LL (scaled)
\item SL, LL (unscaled)
\item SL (scaled)
\item SL (unscaled)
\item LL (scaled)
\item LL (unscaled)
\item SL Order 1 Nod 1
\item SL Order 1 Nod 2
\item SL Order 2 Nod 1 plus bonus order
\item SL Order 2 Nod 2 plus bonus order
\item LL Order 1 Nod 1
\item LL Order 1 Nod 2
\item LL Order 2 Nod 2 plus bonus order
\item LL Order 2 Nod 1 plus bonus order
\item SH, LH (scaled) [archival data only]
\item SH, LH (unscaled) [archival data only]
\item SL, SH (scaled) [archival data only]
\item SL, SH (unscaled) [archival data only]
\end{description}
In most cases, the end user will require just the SL, LL (scaled)
version of the spectrum or equivalent high-resolution spectrum.
\\
\textbf{pipelineVersion}: Spitzer data reduction pipeline version.\\
\textbf{objectName}: Chosen name for object, usually the oldest historically.\\
\textbf{alternateName}: Other useful identifiers.\\
\textbf{quality}: Data quality notes are: 'OK'; 'Wa' (warning, use
spectrum with caution); 'Fa' (fatal, spectrum is bad) -- see the notes
field for an explanation [archival data only]\\
\textbf{notes}: quality control comments detailing any data complexities [archival data only]\\

Spectra are then available seperately as single files or a large
archive, with or without additional photometry points. Please see the
README file for further information.

\subsection{MIPS-SED point source data}

The MIPS-SED data are available as separate files with incorporated
metadata. Column headings are:

\noindent\textbf{SAGESpecID}: This is a unique identifier for each
point source. MIPS-SED sources are labelled with numbers 500--562,
which are obtained by ordering the sources sequentially according to
Right Ascension (R.A.).\\ \textbf{label}: This is a unique identifier
for each spectrum. It is constructed by concatenating R.A. (to 4
decimal places) and declination (dec; to 4 decimal places), instrument
name (MIPSSED) and an optional integer increment for sources with more
than one spectrum, for example: 85.1799-70.1862\_MIPSSED\_2.\\ 
\textbf{ra}: Right ascension co-ordinates (J2000)\\ \textbf{dec}:
Declination co-ordinates (J2000)\\ \textbf{objectName}: Chosen name
for object.\\ \textbf{quality}: This can be one of \textquotedblleft
good\textquotedblright, \textquotedblleft ok\textquotedblright,
\textquotedblleft poor\textquotedblright\ and \textquotedblleft
bad\textquotedblleft. These terms are explained in
Section~\ref{MIPS-QC}.\\ \textbf{waveLength}: wavelength in
microns.\\ \textbf{flux}: flux in Jansky units.\\ \textbf{fluxUnc}:
flux uncertainty in Jansky units.\\ \textbf{SN}: signal-to-noise
ratio.\\ \textbf{sky}: sky noise.

\subsection{IRS \& MIPS-SED extended region data}

Metadata for extended region data comprises of:

\noindent\textbf{Name}: Chosen name for object; SSDR represents SAGE-Spec
Diffuse Region, DEM numbers are taken from the catalogue of
\citet{dav76}.\\
 \textbf{ra}: Right ascension co-ordinates
(J2000)\\ 
\textbf{dec}: Declination
co-ordinates (J2000)\\
\textbf{IRS SL AORID}: AOR number for IRS SL observation.\\
\textbf{IRS LL AORID}: AOR number for IRS LL observation.\\
\textbf{MIPS-SED AORID}: AOR number for MIPS-SED observation.\\

\noindent This data is also presented in Table~\ref{extregions}.

\section[Format of plots]{Format of the plots which accompany the data delivery}

\begin{centering}
\begin{figure}
\includegraphics[angle=90,width=13cm]{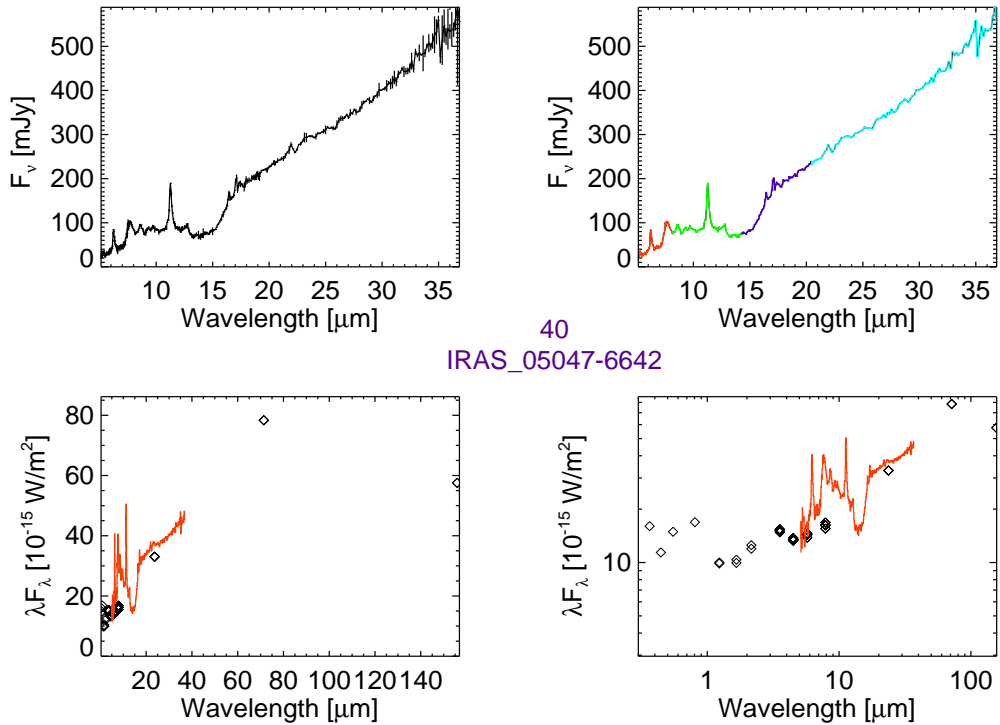}
\caption{An example plot of the IRS staring data for SAGE-Spec source
  40 (see text).}
\label{PlotExample}
\end{figure}
\end{centering}

Each plot, which has a filename of {\tt <label>.ps} (e.g.,
76.2155-66.6354\_IRSX\_SL\_LL\_scaled.ps), is composed of four panels
(see Fig.~\ref{PlotExample}). Clockwise from the top-left these panels
represent:
\begin{description}
\item Full IRS spectrum plotted with error bars.
\item Full IRS spectrum plotted with modules in different colours to show where the joins lie.
\item Spectral energy distribution with photometric points and IRS spectrum, shown in log scale.
\item Spectral energy distribution with photometric points and IRS spectrum, shown in linear scale.
\end{description}
Each plot also includes the SAGE-Spec ID and the chosen point source
name.

\section{Photometric matching of point source data}

For each SAGE-Spec and archival source a cone search using a radius of
20\arcsec\ is performed against the SAGE-LMC photometric database. The
possible matches are retained in the following manner:

\begin{enumerate}
\item Only the closest match in each of the categories below are chosen.
\item Each of these closest matches is further restricted:
\begin{enumerate}
  \item IRAC (3.6, 4.5, 5.8, 8.0\,$\mu$m) matches must be within 3\arcsec.
  \item MIPS 24$\mu$m matches must be within 3\arcsec.
  \item MIPS 70$\mu$m matches must be within 10\arcsec.
  \item MIPS 160$\mu$m matches must be within 20\arcsec.
  \item Two Micron All Sky Survey (2MASS) JHK matches must be within
    3\arcsec.
  \item Infrared Survey Facility (IRSF) JHK matches must be within
    3\arcsec.
  \item Magellanic Clouds Photometric Survey (MCPS) UVBI matches must
    be within 3\arcsec.
\end{enumerate}
\end{enumerate}

\section{Core team members}
\label{sec:coreteam}
\textbf{Ciska Kemper (Academia Sinica, Institute of Astronomy and 
Astrophysics/University of Manchester)}: PI, science overview, observing strategy.\\
\textbf{Karl Gordon (Space Telescope Science Institute)}:
IRS extended region and MIPS-SED data reduction.\\
\textbf{Margaret Meixner (Space Telescope Science Institute)}: science
overview, observing strategy, database, SAGE-LMC photometry liaison.\\
\textbf{Remy Indebetouw (University of Virginia/NRAO)}: quality
control, IRS extended region data reduction.\\
\textbf{Massimo Marengo (Iowa State University)}: source
classification.\\
\textbf{Greg Sloan (Cornell University)}: IRS staring data reduction,
observing strategy.\\
\textbf{Xander Tielens (Leiden Observatory)}: science overview. \\
\textbf{Jacco van Loon (Keele University)}: observing strategy, quality control.\\
\textbf{Paul Woods (University College London/University of Manchester)}:
 Project Manager, quality control, source classification.

\section{Extended team members}
V. Antoniou (Iowa), J.-P. Bernard (CESR, Toulouse), R. D. Blum (NOAO),
M. L. Boyer (STScI), C.-H. R. Chen (U. Virginia), M. Cohen (Berkeley),
C. Dijkstra, M. Galametz (CEA, Saclay), F. Galliano (CEA, Saclay),
C. Gielen (KU Leuven), V. Gorjian (JPL/Caltech), J. Harris (Steward
Observatory), H. Hirashita (ASIAA), S. Hony (CEA, Saclay), J. Hora
(Harvard CfA), R. Indebetouw (U. Virginia/NRAO), O. Jones
(U. Manchester), A. Kawamura (Nagoya University), K. E. Kraemer
(AFRL), E. Lagadec (ESO), K.-H. Law (JHU), B. Lawton (STScI),
J. Leisenring (U. Virginia), S. Madden (CEA, Saclay), M. Matsuura
(UCL), I. McDonald (U. Manchester), C. McGuire (U. Manchester),
B. O'Halloran (Imperial College London), K. Olsen (NOAO), J. Oliveira
(Keele), M. Otsuka (STScI), R. Paladini (SSC), D. Paradis (SSC),
W. T. Reach (SSC), D. Riebel (StScI), D. Rubin (CEA, Saclay),
K. Sandstrom (MPIA), B. Sargent (STScI), J. Seale (STScI), M. Sewi\l o
(STScI), B. Shiao (STScI), A. K. Speck (U. Missouri), S. Srinivasan
(IAP), R. Szczerba (N. Copernicus Astronomical Center. Toru\'n),
A. G. G. M. Tielens (Leiden Observatory), E. van Aarle (KU Leuven),
J. van Benthem (Cornell), S. D. Van Dyk (SSC), H. Van Winckel (KU
Leuven), U. Vijh (U. Toledo), K. Volk (STScI), B. Whitney (Space
Science Institute, Boulder), A. Wilkins (Cornell), M. Wolfire
(U. Maryland), R. Wu (CEA, Saclay), A. Zijlstra (U. Manchester).

\section[Contact information]{Contact information}

\textbf{General questions}: Ciska Kemper
(ciskakemper@fastmail.net), Paul Woods
\\(pmw@star.ucl.ac.uk)\\
\textbf{MIPS-SED extended data}: Karl Gordon (kgordon@stsci.edu)\\
\textbf{IRS extended region data}: Karl Gordon
(kgordon@stsci.edu), Remy Indebetouw (remy@virginia.edu) \\
\textbf{MIPS-SED point sources}: Jacco van Loon (jacco@astro.keele.ac.uk)\\
\textbf{IRS staring data}: Greg Sloan
(sloan@isc.astro.cornell.edu), Paul Woods (pmw@star.ucl.ac.uk)\\
\textbf{Database questions}: Bernie Shiao (shiao@stsci.edu), Margaret
Meixner (meixner@stsci.edu)

\section[Acknowledgments]{Acknowledgments}

This publication makes use of data products from the Two Micron All
Sky Survey (2MASS), which is a joint project of the University of
Massachusetts and the Infrared Processing and Analysis
Center/California Institute of Technology, funded by the National
Aeronautics and Space Administration and the National Science
Foundation.

\noindent MCPS data are publicly available from\\
\url{http://ngala.as.arizona.edu/dennis/mcsurvey.html}

\noindent The IRSF website is at:
\url{http://www.z.phys.nagoya-u.ac.jp/~irsf/index_e.html}

\chapter[IRS spectroscopy]{IRS spectroscopy}

\section{Point sources}

\subsection{Observation settings}

All IRS observations of point sources in this data delivery were
obtained in staring mode. The majority used SL, usually in combination
with LL, and we have delivered these data as full low-resolution
spectra (lores; 5--37$\mu$m).  A minority used the Short-High (SH)
module, usually with Long-High (LH), and we have delivered these where
possible as full high-resolution spectra (hires; 10--37\,$\mu$m). Some
observers used a combination of SL, SH and LH; we have delivered these
data in a SL-hires combination (slhi).

Each target in the SAGE-Spec program was observed after a
high-accuracy peak-up, usually on the target itself.  In cases where
the target was too bright or faint or in a crowded region, the
telescope peaked up on a nearby 2MASS source. Most of the spectra
observed in other programs also used a similar strategy, but some used
moderate-accuracy peak-up, blind peak-up, or the PCRS camera on
Spitzer.

In the SAGE-Spec program, observations were designed so that the
number and length of integration cycles matched for the two
apertures matched in low-resolution modules, in order to maximize
the options for background subtraction.  Most of the larger programs
designed by other users followed a similar course, but not all.  The
user of these data should keep in mind that with the large number of
different astronomers designing observations, virtually any combination
of apertures, integration times, and number of cycles is possible.
As an example, one program observed using only the SL order-1 aperture.

All SAGE-Spec IRS staring data were obtained by the Spitzer Space
Telescope from IRS Campaigns 47 to 53.  These covered the period from
January 9 to August 16 2008. The SAGE-Spec data delivery includes IRS
observations obtained by other programs beginning with in-orbit
checkout (IOC) and continuing all the way through Cycle 5.

\subsection{Data processing}
\label{IRS-DataProc}

The IRS data in this delivery were processed using the SSC pipeline
S18.18.  Before spectra were extracted, the background was subtracted
from each image, and the resulting difference image was cleaned to
replace pixels flagged as bad with data from neighbouring pixels.
Spectra were extracted from individual images, then calibrated and
co-added.

\subsection{Low-resolution spectra}

For SL, the default background subtraction was an aperture difference,
where an image with the source in one aperture served as the
background for the image with the source in the other aperture, but
the same nod position.  For example, for SL1 nod 1, the background
image was SL2 nod 1.  For LL, the default was a nod difference, where
the image with the source in one nod position served as the background
for the image with the source in the other nod position in same
aperture.  Thus LL1 nods 1 and 2 served as backgrounds for each other,
and likewise for LL2 nods 1 and 2.

In many cases, complex backgrounds or additional sources in the
on-target aperture or off-target aperture forced us to modify the
background subtraction.  In addition to aperture or nod differences,
we occasionally resorted to cross differences (other aperture, other
nod).  In some cases, no suitable background subtraction was possible,
which could result in compromised or useless spectroscopy. See
Section~\ref{IRS-QC} for more detailed information.

We relied on the {\tt bmask} images supplied with each data image by
the SSC and the campaign rogue masked produced by the SSC to identify
bad pixels.  Any pixel with a {\tt bmask} value of 4096 or higher was
replaced.  Any pixel which was identified as a rogue in any two
campaigns from the start of the mission through Camp. 54 was also
replaced.  Additionally, all NaNs in the data were replaced.

To replace bad pixels, we applied the {\tt imclean.pro} IDL procedure
written at Cornell University.  This algorithm now serves as the core
of the pixel-replacement within {\tt irsclean.pro}, which is provided
by the SSC.  To replace a bad pixel, the spatial profile of
neighbouring rows are scaled to the row with the bad pixel.  A
replacement value is constructed from good data in the same column.
This algorithm assumes that spatial structure of the source is
unchanged over the rows in question and thus it could break down in
the vicinity of spectral changes in spatial structure, such as
spatially extended emission from a forbidden line near an unresolved
continuum source.

To extract spectra from the spectral images, we used the profile,
ridge, and extract modules of the SSC pipeline.  These are available
within SPICE.  The resulting extraction is equivalent to a
tapered-column extraction within SMART, with the extraction aperture
two pixels wide in the dispersion direction and increasing in width in
the cross-dispersion (spatial) direction linearly with wavelength.
The extraction apertures are defined as follows:
\begin{center}
\begin{tabular}{lcc}
order    & width (pix) & wavelength ($\mu$m)\\
SL2      & 4.000 & 6.0 \\
SL-bonus & 5.333 & 8.0 \\
SL1      & 8.000 & 12.0 \\
LL2      & 4.250 & 16.0 \\
LL-bonus & 5.445 & 20.5 \\
LL1      & 7.172 & 27.0 
\end{tabular}
\end{center}

Spectra are extracted from each spectral image in a given nod
position, then calibrated from digital units to flux density units
(Janskys).  The spectral correction is based on comparison of
similarly processed spectra of calibration stars to spectral templates
for those stars produced at Cornell. HR 6348 (K0 III) served as the
calibrator for all SL data.  It and the K5 giant HD 173511 (K5 III)
served as the calibrators for all LL data.

The \textquotedblleft truth\textquotedblright\ spectra of these
standards were constructed using ratios to low-resolution IRS spectra
of the A1 dwarfs $\alpha$ Lac and $\delta$ UMi, using a generic Kurucz
model of an A1 dwarf as the starting \textquotedblleft
truth\textquotedblright\ spectrum. This new calibration differs from
previous deliveries because we have dropped our assumed photometric
levels by 5\% to bring them into line with the current 24-$\mu$m MIPS
calibration \citep{eng07,rie08}. A paper describing this process in
more detail is in preparation (Sloan et al. 2011).

The spectrum reported for each nod position includes the mean of the
individually extracted spectra in that position, and the formal
uncertainty in the mean (standard deviation / square root of the
number of images).  The individual nod spectra are included in this
data delivery.

To produce a complete spectrum for SL and LL, which are also included
in this delivery, we average the spectra from the two nods, and we
append the data from the three orders to each other.  When combining
the nods, spikes and divots which appear in only one nod but not the
other are rejected.  In these cases, only the data from the
better-behaved nod are used.  New uncertainties in the mean are
calculated (before spike rejection, not after), and if they are larger
than the propagated error from the two nods, are used in their place.
Note that the formal uncertainty in the mean of two data is just half
the difference in the data.

Thus this data delivery includes five spectra from SL and five from LL
(the combined spectrum from the module and the four components from
both nods in both apertures).  In addition, we also provide a combined
low-resolution spectrum, produced by appending the LL data to the SL,
and a final low-resolution spectrum, corrected for discontinuities
between segments (stitched), and with extraneous data removed
(trimmed).

To produce this final spectrum, we assume that the discontinuities
between spectral segments arise from pointing errors, which lead to
partial truncation of mispointed sources by the slit edges.  Thus,
segments are calibrated upwards to the presumably best-centered
segment.  This usually means that SL is scaled upward to LL, since
pointing errors have more impact in the smaller SL slit.  The
corrections are multiplicative scalar corrections and have no
dependence on wavelength.  This assumption is reasonable for
corrections of roughly 10\% or less.  For larger corrections, the
general shape of the continuum can no longer be relied upon as
accurate.  The bonus-order data have been combined with overlapping
first- and second-order data before spectra are corrected for
discontinuities.  The corrections are calculated using the wavelengths
of overlap between adjacent spectral segments.  In the case of SL2 and
LL2, the bonus order which were taken at the same time, and with the
target at the same place in the aperture, provide the overlap needed
between first- and second-order data.

Two more modifications have been made to spectra in the final
delivered product.  Any uncertainty which suggests a signal/noise
ratio greater than 500 is reset so that the S/N equals 500.  Finally,
spectra are truncated to delete data at the ends of the orders which
cannot be calibrated, using the following wavelength ranges:
\begin{center}
\begin{tabular}{lc}
order    & wavelength range ($\mu$m)\\
SL2      & 5.10 -- 7.59 \\
SL-bonus & 7.73 -- 8.39 \\
SL1      & 7.59 -- 14.33 \\
LL2      & 14.20 -- 20.54 \\
LL-bonus & 19.28 -- 21.23 \\
LL1      & 20.46 -- 37.00
\end{tabular}
\end{center}

Note that from 20.46 to 20.54\,$\mu$m, all three LL segments are
valid.  The regions of overlap between segments are the regions used
to determine corrections for discontinuity.

\subsection{High-resolution spectra}

This data delivery also contains spectra observed with the SH and LH
modules.  In these modules, the nod positions are close together
compared to the size of the point-spread function (PSF), forcing us to
use dedicated background observations, matching in the number of
cycles and integration time, for background subtraction. However,
these backgrounds were generally not obtained until about two years
into the Spitzer mission.  Consequently, for many of these early
observations, we must skip the background-subtraction step.

From this point forward, the data processing follows a similar
sequence as for SL and LL. Images were cleaned of pixels with NaNs,
flagged as rogues, or identified as otherwise flawed using {\tt
imclean.pro}.  Spectra were extracted using a full-slit extraction,
which integrates all of the signal in the slit.

The spectra were calibrated using $\xi$ Dra (K2 III) as a standard.
The assumed \textquotedblleft truth\textquotedblright\ spectrum of
$\xi$ Dra is based on a model by
\citet{dec04}\footnote{\url{http://irsa.ipac.caltech.edu/data/SPITZER/docs/irs/calibrationfiles/decinmodels/}},
but with some modifications. First, the OH band strength was increased
by 60\% to reflect the strength as observed by the IRS in LL2. Second,
the overall photometry was dropped by 3\% to match the low-resolution
photometric calibration (and the MIPS 24\,$\mu$m
calibration). Finally, we have shifted the shape of the spectrum by
1\% from 15.5 to 10\,$\mu$m so that the SH data agree with the spectra
of the same sources in SL. As before, Sloan et al.\ (2011) are
preparing a more thorough explanation of these steps.

As with the low-resolution modules, the final step in the data
reduction is to stitch the segments together and trim the extraneous
wavelengths.  Note that SH was stitched to LH, but no orders were
adjusted relative to other orders within the same module, because they
were obtained at the same time with identical telescope pointing.  The
wavelength ranges adopted are as follows:

\begin{center}
\begin{tabular}{lcc}
 & SH wavelength & LH wavelength\\
order & range ($\mu$m) & range ($\mu$m)\\
20 &   9.87--10.40  &  19.06--19.78 \\
19 &  10.40--10.96  &  19.78--20.91 \\
18 &  10.96--11.60  &  20.91--22.11 \\
17 &  11.60--12.32  &  22.11--23.48 \\
16 &  12.32--13.12  &  23.48--25.10 \\
15 &  13.12--14.08  &  25.10--26.79 \\
14 &  14.08--15.09  &  26.79--28.83 \\
13 &  15.09--16.32  &  28.83--31.45 \\
12 &  16.32--17.77  &  31.45--34.07 \\
11 &  17.77--19.47  &  34.07--36.92 \\
\end{tabular}
\end{center}

\subsection{Quality control of the data}
\label{IRS-QC}

The Basic Calibrated Data (BCD) generated by the Spitzer pipeline was
checked visually by members of the SAGE-Spec team. Various data errors
were flagged, and BCDs which would require deviation from the standard
processing (Sect.~\ref{IRS-DataProc}) were noted. Errors found
included FUDL (Fast Uplink Downlink) errors, caused by improper data
transmission from the spacecraft, cosmic ray hits on the CCD, spectral
images which contained more than one spectral trace (i.e., detection
of a nearby source in addition to the targeted source) and
\textquotedblleft jailbarring\textquotedblright, where repeating
vertical patterns (residuals) every four columns are left by Spitzer's
A/D converters in cases of a high background on the peak-up arrays
(e.g., if the emission from a bright source happens to fall on a
peak-up array). In addition, background emission or gradients may have
required a change in the standard subtraction algorithm, as detailed
above (Sect.~\ref{IRS-DataProc}). One DCE (Data Collection Event) was
visually examined per EXPID (exposure ID), of which there were usually
several per module in each BCD. In general, these errors were
corrected by:\\
\textbf{FUDL errors}: The spectra from the affected images were
dropped from further consideration.  In addition, images with FUDL
errors were not used for background subtraction.\\
\textbf{cosmic rays}: Most cosmic ray hits were flagged as such and
replaced by our {\tt imclean.pro} procedure.  Images with unflagged
cosmic ray hits were treated in the same way as images with FUDL
errors if the affected pixels were in the vicinity of the source (or
the source in another image if the affected image was used as a
background).\\
\textbf{multiple sources and background issues}: These generally
required that we modify the background subtraction scheme.  In some
cases, we were forced to drop one of the nod spectra from
consideration because there was no way to avoid problems caused by
other sources or background gradients (either in the target image or
in a background image).  In a few cases, we were unable to mitigate
for the problems, and these are noted in Table B.1.\\

\subsection{Caveats}

Users of the staring-mode IRS spectra should be sure to examine the
error bars to the spectra because these they indicate confidence
limits and potential problems with the data.  In particular, large
error bars throughout a spectral segment indicate that the two spectra
from the individual nod positions do not agree, probably due to
additional sources or background gradients in the slit which we were
unable to correct.  Also, a large error bar for a single pixel when
surrounded by much smaller error bars indicates that at that
wavelength, one of the two nod spectra contained a large spike or
divot that was rejected when the nods were combined into a single
spectrum.  The error bar was calculated based on the rejected datum
and retained.

Users should also be aware that the LL slit is significantly larger
than the SL slit (10\arcsec across compared to 3.6\arcsec), and that
many of the SAGE-Spec sources show some spatial structure.  Often, the
SL segments were normalized up to the LL segments more than the
$\sim$10--15\% expected for random pointing errors for unresolved
sources.  The CORRSL2, CORRSL1, CORRLL2, and CORRLL1 keywords in the
spectral file headers contain the factors by which each segment was
divided to align the spectra; anything smaller than 0.85 indicates a
spatially extended source.

The large LL slit often contains excess background emission, which will
result in the spectrum beginning to rise in the LL1 spectral segment.
This rise to the red is often, but not always, accompanied by larger
error bars, since the spectra from the sky subtraction will have differing
degrees of success in the two nods in these cases.

\subsection{Point source classification}

All IRS spectra of point sources will be classified according to
object type and spectral type. In delivery \#3 we provide the object
classification for the 197 original SAGE-Spec sources, and the
remainder will be delivered in 2011. Point sources are classified as
follows: low mass ($M<8$\,M$_\odot$) post-Main Sequence stars are
classified by chemistry (O- or C-rich) and by evolutionary stage
(Asymptotic Giant Branch; post-Asymptotic Giant Branch; and Planetary
Nebula), hence our groupings {\tt O-AGB, O-PAGB, O-PN, C-AGB, C-PAGB,
  C-PN}. {\tt O-PAGB} has a subgroup for RV\,Tauri stars, {\tt RVTAU}.
Red supergiants have a class of their own, {\tt RSG}. Young stellar
objects can be identified as {\tt YSO1}--{\tt YSO3} (increasing in
evolution) and {\tt YSO4} (candidate Herbig AeBe stars). Stars showing
a stellar photosphere, but no additional dust or gas features are
classified as {\tt STAR}. We also distinguish galaxies ({\tt GAL}),
and H{\sc ii} regions ({\tt HII}). Finally, we use the classification
{\tt OTHER} for objects of \emph{known type} which do not fit into
another category (e.g., R Coronae Borealis stars), and also {\tt UNK}
for objects which cannot be classified (unknown objects). For full
details of the classification process, please see \citet{woo11}.

\section{Extended sources}

\begin{center}
\begin{longtable}{lccccc}
\caption{SAGE-Spec IRS Extended Regions} \label{extregions} \\ 
\hline
\textbf{Name} & \textbf{RA(2000)} & \textbf{Dec(2000)} &
    \textbf{IRS SL AORID} & \textbf{IRS LL AORID} & \textbf{MIPS-SED AORID} \\ \hline
\endfirsthead

\multicolumn{6}{c}%
{{\bfseries \tablename\ \thetable{} -- continued from previous page}} \\
\hline
\textbf{Name} & \textbf{RA(2000)} & \textbf{Dec(2000)} &
    \textbf{IRS SL AORID} & \textbf{IRS LL AORID} & \textbf{MIPS-SED AORID} \\ \hline
\endhead

\hline \multicolumn{6}{r}{{continued on next page\ldots}} \\ \hline
\endfoot
\hline \hline
\endlastfoot
SSDR 1    & 83.009083    &  -68.470444 & 22460928 & 22460928 & 22465024 \\
SSDR 2    & 85.925042    &  -68.252056 & 22461184 & 22461184 & 22465280 \\
SSDR 3    & 78.931834    &  -68.055639 & 22461440 & 22461440 & 22465536 \\
SSDR 4    & 71.920208    &  -67.208611 & 22461696 & 22461696 & 22465792 \\
SSDR 5    & 88.975792    &  -68.199194 & 22461952 & 22461952 & 22466048 \\
SSDR 6    & 86.817875    &  -70.715417 & 22462208 & 22462208 & 22466304 \\
SSDR 7    & 83.789000    &  -70.056722 & 22462464 & 22462464 & 22466560 \\
SSDR 8    & 81.604875    &  -67.485583 & 22462720 & 22462720 & 22466816 \\
SSDR 9    & 83.044708    &  -68.353000 & 22479360 & 22479360 & 22480384 \\
SSDR 10   & 83.095625    &  -66.478194 & 22479872 & 22479872 & 22480896 \\
SSDR 11   & 85.915208    &  -68.771833 & 24241664 & 24241664 & 24242432 \\
SSDR 12   & 82.779583    &  -68.320000 & 24242176 & 24242176 & ---      \\
DEM L 8	  & 73.026292    &  -66.924194 & 22469120 & 22469120 & 22474240 \\
DEM L 10  & 73.049584    &  -69.345306 & 22469376 & 22469376 & 22474496 \\
DEM L 34  & 74.208709    &  -66.413889 & 22469632 & 22469632 & 22474752 \\
DEM L 40  & 74.421709    &  -67.648306 & 22469888 & 22469888 & 22475008 \\
DEM L 55  & 75.421708    &  -70.646889 & 22470144 & 22470144 & 22475264 \\
DEM L 86  & 77.483292    &  -68.900889 & 22470400 & 22470400 & 22475520 \\
DEM L 188 & 81.267917    &  -71.463111 & 22470656 & 22470656 & 22475776 \\
DEM L 243 & 83.879583    &  -66.044111 & 22470912 & 22470912 & 22476032 \\
DEM L 308 & 86.224583    &  -67.348389 & 22471168 & 22471168 & 22476288 \\
DEM L 323 & 87.218292    &  -70.064305 & 22471424 & 22471424 & 22476544 \\
30 Dor	  & 84.7063      &  -69.0993   & 18630144 & 18635008 & 18633728 \\
\hline
\end{longtable}
\end{center}

\subsection{Observation settings}

All the IRS extended source observations in this data delivery were
obtained using the standard IRS spectral mapping template.  The names,
central coordinates, and AOR numbers for the IRS observations are
given in Table~\ref{extregions}.  The regions mapped were chosen to be
large enough to sample the entire region or a representative region
(usually a radial strip).  All the observations were paired with
dedicated background observations taken in a low surface brightness
region located outside of the LMC.

\subsection{Data processing}

The pipeline-reduced IRS spectral mapping data were retrieved from the
Spitzer archive (versions S17.0.4 to S18.7.0).  These individual
observations were combined into spectral cubes using the CUBISM
software \citep{smi07}.  The dedicated off-LMC background observations
were subtracted from the on- and off-source mosaics prior to combining
using CUBISM.  Each extended source spectral mapping observation
results in six different order cubes (except for 30 Dor that only has
four order cubes).  These six different order cubes per object are the
deliverables for this release.  Later releases will include a single
merged spectral cube covering the entire IRS spectral range.  Full
astrometry for each cube (both spatial and spectral) is captured as
part of the FITS header of the spectral cube.

\subsection{Quality control of the data}
\label{IRS-Map-QC}

The quality control on the IRS spectral mapping observations was
focused on examining the mosaics. Extracted IRS spectra frequently
show offsets between orders and modules, with a variety of causes.
Most common for extended sources are differences in what part of the
sky is sampled by different size slits and a wavelength-dependent
telescope beam that is larger than the slit.  As described in the IRS
Instrument Handbook (Chapter 4), the spectra are calibrated for
unresolved point sources.  Not all of the light from such a source is
present in such a spectrum -- some does not fall down the slit (Slit
Loss), and some does not fall in the standard extraction aperture
(Aperture Loss).  For a source with uniform surface brightness on
spatial scales larger than the slit and the telescope beam, light lost
from the part of the sky being measured is compensated by light from
neighboring parts of the sky that get spread into the Slit and
Aperture by the finite spatial response (point spread function or
beam).  The extracted spectrum of such a uniform source must be
divided by the Slit and Aperture Loss Correction Factors, effectively
undoing the corrections required for a point source.  In reality,
extended sources are neither unresolved point sources nor have uniform
surface brightness, so the precise correction to use is unknown.  We
adopt standard practice for IRS extended source software such as
CUBISM, and remove the Slit and Aperture Loss Correction Factors
(i.e., we use the uniform-source calibration). Point or nearly point
sources in the map will have lower flux densities than they should,
and the discrepency will vary by module.  Cubes created with both the
uniform-source and point-source calibrations still have offsets
between orders and modules, especially for fainter regions.

Many investigators calculate and apply a multiplicative correction
between IRS modules, to "order-match" the spectra.  We found that for
faint sources, a purely multiplicative correction could not
universally match all orders, modules, and simultaneously IRAC and
MIPS photometry, but that an additive offset is sometimes present,
which we believe to be largely due to incorrect dark current
subtraction in the SSC pipeline.

The dark current has been observed to vary with time and chip
location.  This is described as \textquotedblleft dark
settle\textquotedblright\ in the IRS Data Reduction Handbook (v4.0.1,
Recipe 11).  That recipe, and SSC-provided software for correcting
dark settle, applies to the LH module, but our analysis and private
communication with SSC suggests that the issue likely exists for all
modules.  If the dark frame used in the BCD pipeline is different from
the actual dark current, an additive offset as a function of chip
location (hence wavelength and spatial position) will be introduced.
More generally, since the SSC has adopted average darks for the entire
mission, variations in the dark during the mission, and any errors in
modelling the time-dependent zodiacal light, will also result in
additive errors in extracted spectra.  The SSC-provided dark-settle
correction software calculates the dark current as a function of
detector row from the pixels between spectral orders.  We have
modified their software to operate on the SL and LL modules, but SL in
particular is challenging to correct because there is very little
inter-order area on the chip.  The regions between the spectral data
and the peak-up imagers (PUI) are often contaminated by stray light
from bright sources in the PUI.  See Section 5.1.16 of the IRS
Instrument Handbook (v4.0) for a description of stray light and a
figure showing the layout of the spectral orders and PUI on the
detector.

Our dark-settle correction code operates on the BCD data with the
following steps:
\begin{enumerate}
\item subtract the median off-source background frame
associated with the AOR (the intent is to correct differences in the
dark current between the on-source and off-source observations, which
is a smaller effect than the total dark current)
\item divide by the flat-field 
\item mask all active parts of the chip (spectral on-order, PUI, and a
  few small manually determined very bad groups of pixels)
\item calculate the resistant mean of the
  unmasked pixels in each row, yielding a 1-dimensional correction as
  a function of row 
\item smooth the correction with a Savitzky-Golay smoothing filter
  \citep{sav64}, requiring that the correction be small at the top and
  bottom of the chip
\item subtract the correction from each row of the 2-dimensional data
\item re-multiply by the flat-field  
\end{enumerate}
Subsequent to the dark-settle correction, spectral cubes are
re-created from this corrected BCD data using CUBISM.

We find that for most of the IRS mapping data in the SAGE-Spec
program, applying the darksettle correction to the BCD frames improves
matching between orders and modules in spectra extracted from
subsequently reconstructed spetral cubes.  However, some of the
faintest maps do not show significant improvement, and some data show
a larger number of hot/dark pixels than in the uncorrected cubes.  The
latter problem arises because the dark-settle operation applies a
background (off-source) subtraction, which tends to make rogue pixels
deviate by less relative to the data's rms, and therefore are harder
for CUBISM's outlier detector algorithm to find and flag.  Due to the
uncertainties, we have decided not to deliver the dark-settle
corrected cubes, but are happy to work with invesitigators who wish to
use them, to determine if they improve a particular scientific
inquiry.  Please contact Remy Indebetouw directly (remy@virginia.edu).

\chapter[MIPS-SED spectroscopy]{MIPS-SED spectroscopy}

\section{Point sources}

\subsection{Observation settings}

All the MIPS-SED point source observations in this data delivery were obtained
using the standard MIPS-SED point source template.  The chop distance
and total exposure time was adjusted based on the nearby sources seen
in the MIPS 70\,$\mu$m image and the measured MIPS 70\,$\mu$m
photometry, respectively.

\subsection{Data processing}

The MIPS-SED data were processed through the MIPS Data Analysis Tool
(DAT, v3.10, 3 Jul 2007).  The details of the MIPS DAT are given by
\citet{gor05}.  The product of the MIPS DAT is both on-source
background subtracted (on-off), on-source only, and off-source only
rectified mosaics combining all the appropriate observations in an
AOR.  For the majority of the sources, the on-source background
subtracted mosaics (on-off) was used as this removes the majority of
the background emission and residual instrumental signatures.

The spectra were extracted from the 2D rectified mosaics.  First, a
spatial profile was created by collapsing the 2D mosaic along the
wavelength axis.  In most cases, the maximum in the spatial profile
identified the source position along the slit.  The extraction was done
using the standard width of 5 pixels with residual background
subtraction done using two regions 3 pixels wide located on either side
of the source.  In the few cases where the off-source chop position was
contaminated by another source, the on-source data was used and the
background subtraction using the two background regions on either side
of the source.  For a few, faint sources, the source position was fixed
to the default location.

The extracted spectra were trimmed to reject all data beyond 95
microns as these wavelengths are contaminated by 2nd order flux.  The
spectra were corrected to infinite aperture and calibrated using the
results given by \citet{lu08}.

\subsection{Quality control of the data}
\label{MIPS-QC}

The quality control on the MIPS-SED observations was focused on
examining the mosaics and spatial profile of the source.  The quality
of a spectrum was deemed as \textquotedblleft
good\textquotedblright\ if there was no source in the off-source
position and the spatial profile was distinctly a point source.  A
source was \textquotedblleft ok\textquotedblright\ if there was a
faint source in the off-source position or if its spatial profile was
not as simple as a single point source, but still enough like a point
source to get a decent extraction.  A source was \textquotedblleft
poor\textquotedblright\ if there was a bright source in the off-source
mosaic or it was very extended.  A source was \textquotedblleft
bad\textquotedblright\ if there was no reasonable source to be
extracted.

\section{Extended sources}

\subsection{Observation settings}

All the MIPS-SED extended source observations in this data delivery
were obtained using the standard MIPS-SED mapping template.  The
names, central coordinates, and AOR numbers for the MIPS-SED
observations are given in Table~\ref{extregions}.  The regions mapped
were chosen to be large enough to sample the entire region or a
representative region (usually a radial strip).  All the observations
were paired with dedicated background observations taken in a low
surface brightness region located outside of the LMC.

\subsection{Data processing}

The MIPS-SED data were processed through the MIPS Data Analysis Tool
(DAT, v3.10, 3 Jul 2007).  The details of the MIPS DAT are given by
\citet{gor05}.  The product of the MIPS DAT is both on-source
background subtracted (on-off), on-source only, and off-source only
rectified mosaics combining all the appropriate observations in an
AOR.

For each spectral map, spectral cubes were created using custom IDL
code (C.\ Engelbracht et al., in prep.) that uses the on-source and
off-source MIPS DAT products to create a 3D spectral cube (which is
basically an image at each wavelength).  The dedicated off-LMC
background observations were subtracted from the on- and off-source
mosaics prior to combining using the custom IDL software.  Full
astrometry for the cube (both spatial and spectral) is captured as
part of the FITS header of the spectral cube.  The extended source
calibration of this code has been checked between similar MIPS SED
mapping and 70\micron\ imaging observations in the SINGS Legacy sample
(C.\ Engelbracht et al., in prep.).

The spectral cubes include data beyond 95 microns, but this data
should be used with extreme caution as these wavelengths are
contaminated by 2nd order flux.

\subsection{Quality control of the data}
\label{MIPS-SED-QC}

The quality control on the MIPS-SED spectral mapping observations was
focused on examining the mosaics and fitted continuum/line strengths.
Reasonable results were obtained.  An additional check was made
between extracted spectra of the entire regions and the equivalent
70\micron\ photometry of the same regions.  The results of this
additional check indicates that the flux calibration of the MIPS-SED
spectral mapping is as expected.

\appendix
\appendixpage
\chapter{Availability of spectra and photometry}

\begin{center}
\begin{longtable}{lccccccc}
\caption{Photometry and spectroscopy for SAGE-Spec sources observed in
  IRS staring mode.$^1$} \label{irsstaring} \\

\hline \multicolumn{1}{c}{\textbf{SAGE-Spec source ID}} & \multicolumn{1}{c}{\textbf{UBVI}} & \multicolumn{1}{c}{\textbf{JHK}} & \multicolumn{1}{c}{\textbf{IRAC}} & \multicolumn{1}{c}{\textbf{MIPS}} & \multicolumn{1}{c}{\textbf{SL}}& \multicolumn{1}{c}{\textbf{LL}}& \multicolumn{1}{c}{\textbf{MIPS-SED}}\\ \hline 
\endfirsthead

\multicolumn{8}{c}%
{{\bfseries \tablename\ \thetable{} -- continued from previous page}} \\
\hline \multicolumn{1}{c}{\textbf{SAGE-Spec source ID}} & \multicolumn{1}{c}{\textbf{UBVI}} & \multicolumn{1}{c}{\textbf{JHK}} & \multicolumn{1}{c}{\textbf{IRAC}} & \multicolumn{1}{c}{\textbf{MIPS}} & \multicolumn{1}{c}{\textbf{SL}}& \multicolumn{1}{c}{\textbf{LL}}& \multicolumn{1}{c}{\textbf{MIPS-SED}}\\ \hline 
\endhead

\hline \multicolumn{8}{r}{{continued on next page\ldots}} \\ \hline
\endfoot

\hline \hline\\ \caption{~$^1$Photometry for SSID17, 24, 125 and 133
  has been altered since delivery \#1. The present table represents
  the best available photometry.}  \endlastfoot

1 &
BVI &
\checkmark &
\checkmark &
24 &
\checkmark &
--- & ---\\\hline
2 &
--- &
\checkmark &
\checkmark &
24, 70 &
\checkmark &
\checkmark & \checkmark\\\hline
3 &
\checkmark &
\checkmark &
\checkmark &
24 &
\checkmark &
\checkmark & ---\\\hline
4 &
\checkmark &
\checkmark &
\checkmark &
24 &
\checkmark &
\checkmark & ---\\\hline
5 &
BVI &
\checkmark &
\checkmark &
24, 70 &
\checkmark &
\checkmark & ---\\\hline
6 &
\checkmark &
\checkmark &
\checkmark &
24 &
\checkmark &
\checkmark & ---\\\hline
7 &
BVI &
\checkmark &
\checkmark &
24 &
\checkmark &
\checkmark & ---\\\hline
8 &
BVI &
\checkmark &
\checkmark &
24 &
\checkmark &
\checkmark & ---\\\hline
9 &
--- &
--- &
\checkmark &
24 &
\checkmark &
\checkmark & ---\\\hline
10 &
BVI &
\checkmark &
\checkmark &
24, 70 &
\checkmark &
\checkmark & ---\\\hline
11 &
\checkmark &
\checkmark &
\checkmark &
\checkmark &
\checkmark &
\checkmark & ---\\\hline
12 &
BVI &
\checkmark &
\checkmark &
24 &
\checkmark &
--- & ---\\\hline
13 &
\checkmark &
\checkmark &
\checkmark &
24 &
\checkmark &
--- & ---\\\hline
14 &
UBV &
\checkmark &
\checkmark &
\checkmark &
\checkmark &
\checkmark & ---\\\hline
15 &
--- &
\checkmark &
\checkmark &
24 &
\checkmark &
--- & ---\\\hline
16 &
UBV &
\checkmark &
\checkmark &
70 &
\checkmark &
--- & ---\\\hline
17 &
--- &
--- &
--- &
24 &
---
 &
\checkmark & \checkmark\\\hline
18 &
--- &
--- &
\checkmark &
24 &
\checkmark &
\checkmark & ---\\\hline
19 &
--- &
\checkmark &
\checkmark &
24 &
\checkmark &
--- & ---\\\hline
20 &
\checkmark &
\checkmark &
\checkmark &
24, 160 &
\checkmark &
\checkmark & ---\\\hline
21 &
\checkmark &
\checkmark &
\checkmark &
24 &
\checkmark &
\checkmark & ---\\\hline
22 &
\checkmark &
\checkmark &
\checkmark &
24 &
\checkmark &
\checkmark & ---\\\hline
23 &
\checkmark &
\checkmark &
\checkmark &
24 &
\checkmark &
--- & ---\\\hline
24 &
--- &
--- &
--- &
--- &
\checkmark &
--- & ---\\\hline
25 &
\checkmark &
\checkmark &
\checkmark &
24, 70 &
\checkmark &
\checkmark & ---\\\hline
26 &
\checkmark &
\checkmark &
\checkmark &
24 &
\checkmark &
\checkmark & ---\\\hline
27 &
\checkmark &
\checkmark &
\checkmark &
24 &
\checkmark &
\checkmark & ---\\\hline
28 &
\checkmark &
\checkmark &
\checkmark &
24 &
\checkmark &
\checkmark & ---\\\hline
29 &
\checkmark &
\checkmark &
\checkmark &
24 &
\checkmark &
\checkmark & ---\\\hline
30 &
\checkmark &
\checkmark &
\checkmark &
24 &
\checkmark &
--- & ---\\\hline
31 &
\checkmark &
\checkmark &
\checkmark &
24 &
\checkmark &
--- & ---\\\hline
32 &
\checkmark &
\checkmark &
\checkmark &
24 &
\checkmark &
--- & ---\\\hline
33 &
\checkmark &
\checkmark &
\checkmark &
24 &
\checkmark &
--- & ---\\\hline
34 &
\checkmark &
\checkmark &
\checkmark &
\checkmark &
\checkmark &
\checkmark & ---\\\hline
35 &
\checkmark &
\checkmark &
\checkmark &
24 &
\checkmark &
--- & ---\\\hline
36 &
\checkmark &
\checkmark &
\checkmark &
24 &
\checkmark &
--- & ---\\\hline
37 &
\checkmark &
\checkmark &
\checkmark &
24 &
\checkmark &
--- & ---\\\hline
38 &
\checkmark &
\checkmark &
\checkmark &
24 &
\checkmark &
\checkmark & ---\\\hline
39 &
\checkmark &
\checkmark &
\checkmark &
24 &
\checkmark &
\checkmark & ---\\\hline
40 &
\checkmark &
\checkmark &
\checkmark &
\checkmark &
\checkmark &
\checkmark & ---\\\hline
41 &
UBV &
\checkmark &
\checkmark &
24 &
\checkmark &
--- & ---\\\hline
42 &
\checkmark &
\checkmark &
\checkmark &
24 &
\checkmark &
\checkmark & ---\\\hline
43 &
\checkmark &
\checkmark &
\checkmark &
24 &
\checkmark &
--- & ---\\\hline
44 &
BVI &
\checkmark &
\checkmark &
24, 70 &
\checkmark &
\checkmark & ---\\\hline
45 &
--- &
\checkmark &
\checkmark &
24 &
\checkmark &
--- & ---\\\hline
46 &
--- &
\checkmark &
\checkmark &
24 &
\checkmark &
--- & ---\\\hline
47 &
--- &
\checkmark &
\checkmark &
24 &
\checkmark &
--- & ---\\\hline
48 &
--- &
\checkmark &
\checkmark &
24 &
\checkmark &
--- & ---\\\hline
49 &
--- &
\checkmark &
\checkmark &
24 &
\checkmark &
--- & ---\\\hline
50 &
\checkmark &
\checkmark &
\checkmark &
24 &
\checkmark &
\checkmark & ---\\\hline
51 &
--- &
\checkmark &
\checkmark &
24 &
\checkmark &
\checkmark & ---\\\hline
52 &
--- &
\checkmark &
\checkmark &
24, 70 &
\checkmark &
\checkmark & ---\\\hline
53 &
BV &
\checkmark &
\checkmark &
24, 70 &
\checkmark &
\checkmark & ---\\\hline
54 &
BVI &
\checkmark &
\checkmark &
24 &
\checkmark &
\checkmark & ---\\\hline
55 &
BVI &
\checkmark &
\checkmark &
24 &
\checkmark &
\checkmark & ---\\\hline
56 &
\checkmark &
\checkmark &
\checkmark &
24 &
\checkmark &
\checkmark & ---\\\hline
57 &
\checkmark &
\checkmark &
\checkmark &
24 &
\checkmark &
--- & ---\\\hline
58 &
\checkmark &
\checkmark &
\checkmark &
24 &
\checkmark &
--- & ---\\\hline
59 &
BV &
\checkmark &
4.5, 5.8, 8.0 &
24 &
\checkmark &
--- & ---\\\hline
60 &
\checkmark &
\checkmark &
\checkmark &
24 &
\checkmark &
\checkmark & ---\\\hline
61 &
\checkmark &
\checkmark &
\checkmark &
24 &
\checkmark &
\checkmark & ---\\\hline
62 &
\checkmark &
\checkmark &
\checkmark &
24, 70 &
\checkmark &
\checkmark & ---\\\hline
63 &
BVI &
\checkmark &
\checkmark &
24 &
\checkmark &
--- & ---\\\hline
64 &
\checkmark &
\checkmark &
\checkmark &
24 &
\checkmark &
\checkmark & ---\\\hline
65 &
--- &
\checkmark &
\checkmark &
24 &
\checkmark &
\checkmark & ---\\\hline
66 &
BVI &
\checkmark &
\checkmark &
24 &
\checkmark &
\checkmark & ---\\\hline
67 &
\checkmark &
\checkmark &
\checkmark &
24 &
\checkmark &
--- & ---\\\hline
68 &
UBV &
\checkmark &
\checkmark &
24 &
\checkmark &
--- & ---\\\hline
69 &
\checkmark &
\checkmark &
\checkmark &
24, 70 &
\checkmark &
\checkmark & ---\\\hline
70 &
--- &
K &
4.5, 5.8, 8.0 &
24 &
\checkmark &
\checkmark & ---\\\hline
71 &
--- &
\checkmark &
\checkmark &
24 &
\checkmark &
\checkmark & ---\\\hline
72 &
\checkmark &
\checkmark &
\checkmark &
24 &
\checkmark &
--- & ---\\\hline
73 &
\checkmark &
\checkmark &
\checkmark &
24 &
\checkmark &
\checkmark & ---\\\hline
74 &
\checkmark &
\checkmark &
\checkmark &
24, 70 &
\checkmark &
\checkmark & ---\\\hline
75 &
\checkmark &
\checkmark &
\checkmark &
24 &
\checkmark &
\checkmark & ---\\\hline
76 &
\checkmark &
\checkmark &
\checkmark &
24 &
\checkmark &
--- & ---\\\hline
77 &
\checkmark &
\checkmark &
\checkmark &
24 &
\checkmark &
--- & ---\\\hline
78 &
BVI &
\checkmark &
\checkmark &
\checkmark &
\checkmark &
\checkmark & ---\\\hline
79 &
BVI &
\checkmark &
\checkmark &
24 &
\checkmark &
--- & ---\\\hline
80 &
\checkmark &
\checkmark &
\checkmark &
24 &
\checkmark &
\checkmark & ---\\\hline
81 &
\checkmark &
\checkmark &
\checkmark &
24 &
\checkmark &
--- & ---\\\hline
82 &
\checkmark &
\checkmark &
\checkmark &
24 &
\checkmark &
\checkmark & ---\\\hline
83 &
BVI &
\checkmark &
\checkmark &
24 &
\checkmark &
--- & ---\\\hline
84 &
\checkmark &
\checkmark &
\checkmark &
24, 70 &
\checkmark &
\checkmark & ---\\\hline
85 &
\checkmark &
\checkmark &
\checkmark &
24 &
\checkmark &
\checkmark & ---\\\hline
86 &
\checkmark &
\checkmark &
\checkmark &
24 &
\checkmark &
--- & ---\\\hline
87 &
--- &
\checkmark &
\checkmark &
24 &
\checkmark &
--- & ---\\\hline
88 &
\checkmark &
\checkmark &
\checkmark &
--- &
\checkmark &
--- & ---\\\hline
89 &
\checkmark &
\checkmark &
\checkmark &
24 &
\checkmark &
--- & ---\\\hline
90 &
\checkmark &
\checkmark &
\checkmark &
24 &
\checkmark &
\checkmark & ---\\\hline
91 &
--- &
\checkmark &
\checkmark &
24 &
\checkmark &
--- & ---\\\hline
92 &
\checkmark &
\checkmark &
\checkmark &
24, 70 &
\checkmark &
\checkmark & ---\\\hline
93 &
\checkmark &
\checkmark &
\checkmark &
24 &
\checkmark &
\checkmark & ---\\\hline
94 &
\checkmark &
\checkmark &
\checkmark &
24 &
\checkmark &
\checkmark & ---\\\hline
95 &
\checkmark &
\checkmark &
\checkmark &
24 &
\checkmark &
\checkmark & ---\\\hline
96 &
\checkmark &
\checkmark &
\checkmark &
24 &
\checkmark &
\checkmark & ---\\\hline
97 &
\checkmark &
\checkmark &
\checkmark &
\checkmark &
\checkmark &
\checkmark & \checkmark\\\hline
98 &
BVI &
\checkmark &
\checkmark &
24 &
\checkmark &
\checkmark & ---\\\hline
99 &
BVI &
\checkmark &
\checkmark &
24 &
\checkmark &
--- & ---\\\hline
100 &
\checkmark &
\checkmark &
\checkmark &
24 &
\checkmark &
\checkmark & ---\\\hline
101 &
BV &
\checkmark &
\checkmark &
\checkmark &
\checkmark &
\checkmark & ---\\\hline
102 &
--- &
\checkmark &
\checkmark &
24, 70 &
\checkmark &
\checkmark & \checkmark\\\hline
103 &
--- &
\checkmark &
\checkmark &
24 &
\checkmark &
\checkmark & ---\\\hline
104 &
UBV &
\checkmark &
\checkmark &
\checkmark &
\checkmark &
\checkmark & ---\\\hline
105 &
BVI &
\checkmark &
\checkmark &
24 &
\checkmark &
\checkmark & ---\\\hline
106 &
\checkmark &
\checkmark &
\checkmark &
24, 70 &
\checkmark &
\checkmark & ---\\\hline
107 &
\checkmark &
\checkmark &
\checkmark &
24 &
\checkmark &
\checkmark & ---\\\hline
108 &
\checkmark &
\checkmark &
\checkmark &
\checkmark &
\checkmark &
\checkmark & ---\\\hline
109 &
\checkmark &
\checkmark &
\checkmark &
24, 70 &
\checkmark &
\checkmark & ---\\\hline
110 &
\checkmark &
\checkmark &
\checkmark &
24 &
\checkmark &
--- & ---\\\hline
111 &
\checkmark &
\checkmark &
\checkmark &
24 &
\checkmark &
\checkmark & ---\\\hline
112 &
UBV &
--- &
\checkmark &
24 &
\checkmark &
--- & ---\\\hline
113 &
\checkmark &
\checkmark &
\checkmark &
24 &
\checkmark &
\checkmark & ---\\\hline
114 &
\checkmark &
\checkmark &
\checkmark &
24, 70 &
\checkmark &
\checkmark & \checkmark\\\hline
115 &
\checkmark &
\checkmark &
\checkmark &
24 &
\checkmark &
--- & ---\\\hline
116 &
\checkmark &
\checkmark &
\checkmark &
24 &
\checkmark &
\checkmark & ---\\\hline
117 &
UBI &
\checkmark &
\checkmark &
24 &
\checkmark &
\checkmark & ---\\\hline
118 &
\checkmark &
\checkmark &
\checkmark &
24 &
\checkmark &
\checkmark & ---\\\hline
119 &
\checkmark &
\checkmark &
\checkmark &
24 &
\checkmark &
\checkmark & ---\\\hline
120 &
\checkmark &
\checkmark &
\checkmark &
24 &
\checkmark &
--- & ---\\\hline
121 &
\checkmark &
\checkmark &
\checkmark &
24, 70 &
\checkmark &
\checkmark & ---\\\hline
122 &
UB &
\checkmark &
\checkmark &
24 &
\checkmark &
\checkmark & ---\\\hline
123 &
UB &
\checkmark &
\checkmark &
24 &
\checkmark &
\checkmark & ---\\\hline
124 &
\checkmark &
\checkmark &
\checkmark &
24 &
\checkmark &
--- & ---\\\hline
125 &
--- &
\checkmark &
\checkmark &
24, 70 &
\checkmark &
\checkmark & ---\\\hline
126 &
\checkmark &
\checkmark &
\checkmark &
24 &
\checkmark &
\checkmark & ---\\\hline
127 &
\checkmark &
\checkmark &
\checkmark &
24 &
\checkmark &
--- & ---\\\hline
128 &
UBV &
\checkmark &
\checkmark &
24 &
\checkmark &
\checkmark & ---\\\hline
129 &
UBV &
\checkmark &
\checkmark &
24 &
\checkmark &
\checkmark & ---\\\hline
130 &
BVI &
\checkmark &
\checkmark &
24 &
\checkmark &
\checkmark & ---\\\hline
131 &
\checkmark &
\checkmark &
\checkmark &
24 &
\checkmark &
\checkmark & ---\\\hline
132 &
--- &
\checkmark &
\checkmark &
24 &
\checkmark &
--- & ---\\\hline
133 &
BVI &
\checkmark &
\checkmark &
24 &
\checkmark &
--- & ---\\\hline
134 &
UBV &
\checkmark &
\checkmark &
24 &
\checkmark &
--- & ---\\\hline
135 &
BV &
\checkmark &
\checkmark &
24 &
\checkmark &
\checkmark & ---\\\hline
136 &
--- &
\checkmark &
\checkmark &
24 &
\checkmark &
--- & ---\\\hline
137 &
\checkmark &
\checkmark &
\checkmark &
24 &
\checkmark &
\checkmark & ---\\\hline
138 &
\checkmark &
\checkmark &
\checkmark &
24, 70 &
\checkmark &
\checkmark & ---\\\hline
139 &
\checkmark &
\checkmark &
\checkmark &
24 &
\checkmark &
--- & ---\\\hline
140 &
BV &
\checkmark &
\checkmark &
24 &
\checkmark &
\checkmark & ---\\\hline
141 &
\checkmark &
\checkmark &
\checkmark &
24 &
\checkmark &
\checkmark & ---\\\hline
142 &
\checkmark &
\checkmark &
\checkmark &
24 &
\checkmark &
--- & ---\\\hline
143 &
\checkmark &
\checkmark &
\checkmark &
24 &
\checkmark &
--- & ---\\\hline
144 &
\checkmark &
\checkmark &
\checkmark &
24, 70 &
\checkmark &
\checkmark & ---\\\hline
145 &
BVI &
\checkmark &
\checkmark &
24 &
\checkmark &
\checkmark & ---\\\hline
146 &
\checkmark &
\checkmark &
\checkmark &
24 &
\checkmark &
\checkmark & ---\\\hline
147 &
UBV &
\checkmark &
\checkmark &
24 &
\checkmark &
\checkmark & ---\\\hline
148 &
\checkmark &
\checkmark &
\checkmark &
24 &
\checkmark &
--- & ---\\\hline
149 &
\checkmark &
\checkmark &
\checkmark &
24 &
\checkmark &
\checkmark & ---\\\hline
150 &
\checkmark &
\checkmark &
\checkmark &
24, 70 &
\checkmark &
\checkmark & ---\\\hline
151 &
\checkmark &
\checkmark &
\checkmark &
24 &
\checkmark &
\checkmark & ---\\\hline
152 &
\checkmark &
\checkmark &
\checkmark &
24 &
\checkmark &
--- & ---\\\hline
153 &
\checkmark &
\checkmark &
\checkmark &
24 &
\checkmark &
--- & ---\\\hline
154 &
\checkmark &
\checkmark &
\checkmark &
24, 70 &
\checkmark &
\checkmark & ---\\\hline
155 &
BVI &
\checkmark &
\checkmark &
24 &
\checkmark &
--- & ---\\\hline
156 &
BVI &
\checkmark &
\checkmark &
24 &
\checkmark &
\checkmark & ---\\\hline
157 &
\checkmark &
\checkmark &
\checkmark &
24 &
\checkmark &
\checkmark & ---\\\hline
158 &
BV &
\checkmark &
\checkmark &
24 &
\checkmark &
\checkmark & ---\\\hline
159 &
\checkmark &
\checkmark &
\checkmark &
24 &
\checkmark &
--- & ---\\\hline
160 &
\checkmark &
\checkmark &
\checkmark &
24 &
\checkmark &
\checkmark & ---\\\hline
161 &
\checkmark &
\checkmark &
\checkmark &
24 &
\checkmark &
--- & ---\\\hline
162 &
\checkmark &
\checkmark &
\checkmark &
24 &
\checkmark &
\checkmark & ---\\\hline
163 &
\checkmark &
\checkmark &
\checkmark &
24, 70 &
\checkmark &
\checkmark & \checkmark\\\hline
164 &
BVI &
HK &
\checkmark &
24, 70 &
\checkmark &
\checkmark & ---\\\hline
165 &
\checkmark &
\checkmark &
\checkmark &
24 &
\checkmark &
\checkmark & ---\\\hline
166 &
\checkmark &
\checkmark &
\checkmark &
24 &
\checkmark &
\checkmark & ---\\\hline
167 &
--- &
--- &
\checkmark &
24 &
\checkmark &
\checkmark & ---\\\hline
168 &
UBV &
\checkmark &
\checkmark &
24, 70 &
\checkmark &
\checkmark & ---\\\hline
169 &
UBV &
\checkmark &
\checkmark &
24 &
\checkmark &
\checkmark & ---\\\hline
170 &
UBV &
\checkmark &
\checkmark &
24 &
\checkmark &
\checkmark & ---\\\hline
171 &
UBV &
\checkmark &
\checkmark &
24 &
\checkmark &
\checkmark & ---\\\hline
172 &
\checkmark &
\checkmark &
\checkmark &
24 &
\checkmark &
\checkmark & ---\\\hline
173 &
\checkmark &
\checkmark &
\checkmark &
24 &
\checkmark &
--- & ---\\\hline
174 &
\checkmark &
\checkmark &
\checkmark &
24 &
\checkmark &
\checkmark & ---\\\hline
175 &
\checkmark &
\checkmark &
\checkmark &
24 &
\checkmark &
\checkmark & ---\\\hline
176 &
BVI &
\checkmark &
\checkmark &
24 &
\checkmark &
--- & ---\\\hline
177 &
\checkmark &
\checkmark &
\checkmark &
24 &
\checkmark &
\checkmark & ---\\\hline
178 &
\checkmark &
\checkmark &
\checkmark &
24 &
\checkmark &
--- & ---\\\hline
179 &
BVI &
\checkmark &
\checkmark &
24 &
\checkmark &
--- & ---\\\hline
180 &
\checkmark &
\checkmark &
\checkmark &
24 &
\checkmark &
\checkmark & ---\\\hline
181 &
\checkmark &
\checkmark &
\checkmark &
24 &
\checkmark &
\checkmark & ---\\\hline
182 &
\checkmark &
\checkmark &
\checkmark &
24 &
\checkmark &
\checkmark & ---\\\hline
183 &
\checkmark &
\checkmark &
\checkmark &
24, 70 &
\checkmark &
\checkmark & ---\\\hline
184 &
BVI &
\checkmark &
\checkmark &
24, 70 &
\checkmark &
\checkmark & ---\\\hline
185 &
--- &
\checkmark &
\checkmark &
24 &
\checkmark &
\checkmark & ---\\\hline
186 &
\checkmark &
\checkmark &
\checkmark &
24 &
\checkmark &
\checkmark & ---\\\hline
187 &
\checkmark &
\checkmark &
\checkmark &
24 &
\checkmark &
\checkmark & ---\\\hline
188 &
\checkmark &
\checkmark &
\checkmark &
24 &
\checkmark &
--- & ---\\\hline
189 &
BVI &
\checkmark &
\checkmark &
24 &
\checkmark &
--- & ---\\\hline
190 &
BVI &
--- &
\checkmark &
24, 70 &
\checkmark &
\checkmark & ---\\\hline
191 &
\checkmark &
\checkmark &
\checkmark &
24 &
\checkmark &
--- & ---\\\hline
192 &
\checkmark &
\checkmark &
\checkmark &
24 &
\checkmark &
\checkmark & ---\\\hline
193 &
\checkmark &
\checkmark &
\checkmark &
24 &
\checkmark &
\checkmark & ---\\\hline
194 &
BVI &
\checkmark &
\checkmark &
24 &
\checkmark &
--- & ---\\\hline
195 &
--- &
\checkmark &
\checkmark &
24 &
\checkmark &
--- & ---\\\hline
196 &
\checkmark &
\checkmark &
\checkmark &
24, 70 &
\checkmark &
\checkmark & ---\\\hline
197 &
\checkmark &
\checkmark &
\checkmark &
24 &
\checkmark &
--- & ---\\\hline
\end{longtable}
\end{center}

\chapter[Quality control results for IRS spectra]{Quality control results for SAGE-Spec IRS spectra}

\begin{center}
{\small
\begin{longtable}{lcccc}
\caption{{\normalsize Data processing summary (see key below table).}} \label{tab:IRS-QC} \\

\hline \multicolumn{1}{c}{\textbf{ID}} & \multicolumn{1}{c}{\textbf{Reduction method}} & \multicolumn{1}{c}{\textbf{SL}} & \multicolumn{1}{c}{\textbf{LL}} & \multicolumn{1}{c}{\textbf{Notes}}\\ \hline 
\endfirsthead

\multicolumn{5}{c}%
{{\bfseries \tablename\ \thetable{} -- continued from previous page}} \\
\hline \multicolumn{1}{c}{\textbf{ID}} & \multicolumn{1}{c}{\textbf{Reduction method}} & \multicolumn{1}{c}{\textbf{SL}} & \multicolumn{1}{c}{\textbf{LL}}& \multicolumn{1}{c}{\textbf{Notes}}\\ \hline 
\endhead

\hline \multicolumn{5}{r}{{continued on next page\ldots}} \\ \hline
\endfoot

\hline \hline
\endlastfoot
1   & default      & ---         &              & \\
2   & default      & ---         & ---          & \\                                  
3   & default      & ---         & ---          & \\                                  
4   & default      & J-Wa        & ---          & \\
5   & default      & ---         & ---          & \\                   
6   & modified     & Ray-Mc-Notes& B-Wa         & SL -- 0300: Ray\\
7   & default      & ---         & N-Wa         &       \\  
8   & default      & ---         & ---          & \\    
9   & default      & ---         & ---          & \\   
10  & modified     & ---         & S-Mc-1Dx,2Da & \\
11  & default      & ---         & ---          & \\           
12  & default      & J-Wa        &              & \\    
13  & modified     & B-Mc-Dn     &              & \\
14  & default      & ---         & ---          & \\        
15  & default      & J-Wa        &              & \\  
16  & default      & ---         &              & \\
17  & modified     &             & ---          & \\
18  & default      & ---         & ---          & \\           
19  & default      & J-Wa        &              & \\             
20  & modified     & X-Mc-Notes  & ---          & SL -- 0200: FUDL, 0400: Dx\\     
21  & default      & ---         & ---          & \\                   
22  & modified     & P-Mc-2Dn    & S-Wa         & \\
23  & default      & ---         &              & \\   
24  & modified     & F-Mi-Dn     &              & \\ 
25  & modified     & B-Mc-Dn     & B-Mc-C2      & \\
26  & default      & J-Wa        & ---          & \\
27  & default      & S-Wa        & ---          & \\
28  & default      & S-Wa        & S-Wa         & \\  
29  & default      & J-Wa        & ---          & \\
30  & default      & J-Wa        &              & \\                    
31  & default      & J-Wa        &              & \\     
32  & default      & ---         &              & \\                                 
33  & default      & J-Wa        &              & \\
34  & modified     & B,X-Mc-Notes& ---          & SL -- 0400: bad pixel block\\
35  & default      & ---         &              & \\                                  
36  & default      & ---         &              & \\    
37  & default      & ---         &              & \\          
38  & modified     & J,Ray-Wa-Notes & ---       & SL -- 0502: Ray\\
39  & modified     & ---         & S-Mc-Da      & \\
40  & default      & ---         & ---          & \\  
41  & default      & ---         &              & \\                   
42  & default      & ---         & ---          & \\           
43  & default      & ---         &              & \\                    
44  & modified     & B-Wa        & S-Mc-1Da     & \\
45  & default      & J-Wa        &              & \\         
46  & default      & ---         &              & \\             
47  & default      & ---         &              & \\   
48  & default      & J-Wa        &              & \\
49  & modified     & B-Mc-Dn     &              & \\
50  & modified     & B-Mc-Dn,Notes& B-Mc-Da     & SL -- 14.2$\mu$m spike \\
51  & modified     & ---         & S-Mc-1C1     & \\
52  & default      & ---         & ---          & \\                                                          
53  & default      & ---         & ---          & \\                                  
54  & default      & J-Wa        & ---          & \\
55  & modified     & ---         & N-Mc-1Da     & \\       
56  & modified     & B-Mc-1Dn    & S-Wa         & \\  
57  & default      & ---         &              & \\        
58  & modified     & X-Mc-Notes  &              & SL -- 0404: bad pixel block\\
59  & default      & S-Wa        &              & \\
60  & modified     & ---         & N-Mc-1Da     & \\
61  & default      & ---         & S-Wa         & \\
62  & default      & ---         & ---          & \\           
63  & default      & ---         &              & \\                              
64  & default      & ---         & ---          & \\        
65  & modified     & ---         & S-Mc-1C1     & \\  
66  & default      & ---         & B-Mc-1Da1,1Dx2 & \\
67  & default      & ---         &              & \\    
68  & default      & ---         &              & \\        
69  & modified     & B-Mi-Dn     & B-Mi-1Da     & \\
70  & default      & E-Wa        & ---          & \\
71  & modified     & ---         & N-Mi-Da      & \\
72  & modified     & S-Mc-Dn     &              & \\
73  & default      & ---         & B-Wa         & \\ 
74  & default      & ---         & ---          & \\           
75  & default      & ---         & S-Wa         & \\       
76  & default      & ---         &              & \\                                  
77  & default      & ---         &              & \\     
78  & default      & J-Wa        & ---          & \\
79  & default      & ---         &              & \\   
80  & modified     & ---         & S-Mc-C1      & \\
81  & default      & ---         &              & \\  
82  & default      & ---         & ---          & \\    
83  & default      & N-Wa        &              & \\
84  & default      & ---         & ---          & \\         
85  & modified     & J-Wa        & S-Mc-1Cn1    & \\       
86  & default      & ---         &              & \\    
87  & modified     & B,J-Mi-1Dn  &              & \\
88  & modified     & B,S-Mi-Dn   &              & \\
89  & default      & S-Wa        &              & \\       
90  & default      & ---         & ---          & \\                                  
91  & modified     & B-Mc-2Dn    &              & \\ 
92  & modified     & X-Mc-Notes  & ---          & SL -- 0402: FUDL, 0202: Dx, 0302: Ray\\
93  & modified     & ---         & B,S-Mc-Da    & \\
94  & default      & ---         & ---          & \\   
95  & modified     & B-Mc-1Dn    & B-Mc-1Da     & \\
96  & default      & ---         & ---          & \\                                  
97  & default      & ---         & ---          & \\
98  & modified     & ---         & S-Mc-Notes   & LL -- 0802: FUDL, 0902: Da \\
99  & default      & ---         &              & \\           
100 & default      & ---         & N-Wa         & \\
101 & default      & ---         & ---          & \\       
102 & default      & ---         & ---          & \\
103 & modified     & ---         & S-Mc-1Da     & \\
104 & modified     & J,Ray-Mc-Notes & ---       & SL -- 0502: Ray\\
105 & modified     & J-Wa        & B-Mi-1Dx,2Da & \\
106 & modified     & ---         & B-Mc-1C1     & \\
107 & default      & ---         & ---          & \\     
108 & modified     & B-Mi-Dn     & S-Mc-Da      & \\
109 & modified     & ---         & S-Mc-1Cn2    & \\
110 & default      & ---         &              & \\    
111 & default      & ---         & ---          & \\   
112 & default      & ---         &              & \\         
113 & modified     & ---         & S-Mc-1Da     & \\
114 & default      & J-Wa        & N-Wa         & \\
115 & default      & ---         &              & \\          
116 & modified     & J,S-Mc-Notes& S-Mc-2Da     & SL -- 0401: Ray, 0202: FUDL, 0402: Dx \\ 
117 & modified     & S-Mc-Notes  & N,S-Fa       & SL -- Ridge constrained\\
118 & default      & ---         & ---          & \\   
119 & modified     & ---         & S-Mi-Da      & \\
120 & default      & ---         &              & \\   
121 & default      & ---         & ---          & \\   
122 & default      & ---         & S-Wa         & \\
123 & modified     & ---         & B-Mc-1Da     & \\
124 & default      & ---         &              & \\   
125 & default      & ---         & ---          & \\  
126 & default      & ---         & ---          & \\           
127 & default      & S-Wa        &              & \\
128 & modified     & ---         & S-Mc-Da      & \\
129 & default      & ---         & ---          & \\   
130 & default      & ---         & ---          & \\    
131 & modified     & J-Wa        & B-Mc-Da      & \\
132 & default      & ---         &              & \\   
133 & default      & ---         &              & \\                                  
134 & default      & ---         &              & \\   
135 & default      & ---         & ---          & \\           
136 & default      & ---         &              & \\   
137 & default      & ---         & B-Wa         & \\
138 & default      & ---         & ---          & \\   
139 & default      & B-Wa        &              & \\
140 & default      & ---         & ---          & \\   
141 & default      & J-Wa        & B,S-Wa       & \\
142 & default      & ---         &              & \\   
143 & default      & ---         &              & \\           
144 & default      & ---         & ---          & \\   
145 & default      & ---         & B,S-Wa       & \\
146 & default      & ---         & N-Wa         & \\
147 & default      & ---         & ---          & \\  
148 & modified     & B-Mc-1Dn    &              & \\
149 & default      & ---         & N-Wa         & \\
150 & default      & ---         & ---          & \\
151 & default      & ---         & ---          & \\                                  
152 & default      & B-Wa        &              & \\
153 & default      & ---         &              & \\    
154 & default      & ---         & ---          & \\            
155 & default      & ---         &              & \\   
156 & modified     & B,S-Mc-Dn   & B,S,N-Wa     & \\
157 & default      & ---         & ---          & \\     
158 & modified     & S-Mc-2Dn    & S-Wa         & \\
159 & default      & ---         &              & \\              
160 & modified     & B,J-Mc-Dn   & N-Fa         & \\
161 & modified     & J,P-Mi-Dn   &              & \\
162 & modified     & N-Wa        & S-Mc-1Da     & \\
163 & default      & ---         & S-Wa         & \\
164 & modified     & B,J-Mi-1Dn  & ---          & \\
165 & modified     & Ray-Wa      & S-Mi-Da      & \\
166 & default      & ---         & ---          & \\          
167 & modified     & ---         & B-Mc-1Cn1    & \\         
168 & default      & ---         & ---          & \\   
169 & modified     & B-Mc-1Dn    & B-Mi-Cn2     & \\
170 & modified     & ---         & S-Mi-1Da     & \\       
171 & modified     & S-Mc-Notes  & S-Mi-Notes   & SL -- 0401: FUDL, 0201: Dx \\
    &              &             &              & LL -- 1Da, 2Dn1, 2Dx2\\
172 & modified     & ---         & S-Mc-Da      & \\
173 & modified     & B-Mc-1Dn    &              & \\
174 & modified     & F-Mi-Dn     & F-Wa         & \\
175 & modified     & X-Mc-Notes  & ---          & SL -- 0301: Ray, not dropped, 0501: Dx\\
176 & default      & ---         &              & \\   
177 & default      & ---         & ---          & \\          
178 & default      & ---         &              & \\   
179 & modified     & B-Mc-1Dn    &              & \\ 
180 & default      & ---         & N-Wa         & \\ 
181 & default      & ---         & ---          & \\                                  
182 & default      & ---         & B,N-Wa       & \\
183 & default      & ---         & ---          & \\      
184 & modified     & ---         & N-Mc-1Da     & \\
185 & default      & ---         & ---          & \\     
186 & default      & J-Wa        & ---          & \\
187 & default      & ---         & ---          & \\             
188 & default      & ---         &              & \\                   
189 & default      & ---         &              & \\        
190 & default      & ---         & ---          & \\  
191 & modified     & B-Mi-Dn     &              & \\
192 & default      & ---         & ---          & \\  
193 & default      & ---         & ---          & \\   
194 & default      & ---         &              & \\
195 & default      & ---         &              & \\               
196 & default      & ---         & ---          & \\            
197 & default      & ---         &              &       
\end{longtable}}
\textsc{Key}: -- 
\begin{tabular}{lp{12.2cm}}
\textbf{Reduction method} & \emph{default} -- as described in
Sect.~\ref{IRS-DataProc}; \emph{modified} -- the default method of
background subtraction, extraction and coadding of spectra was changed
in some way, as detailed in a three-part code, constructed as: {\tt{<problem>-<severity>-<solution>}}\\
\textbf{Problem}  & \emph{J}  -- jailbars in the data.\\
                  & \emph{N}  -- significant difference in appearance between two nods.\\
                  & \emph{FUDL} -- FUDL (data download) error.\\
                  & \emph{Ray} -- cosmic ray strike.\\
                  & \emph{P} -- peak-up related problem.\\
                  & \emph{B} -- background issue.\\
                  & \emph{S} -- additional source.\\
                  & \emph{E} -- excess emission.\\
                  & \emph{F} -- very faint source.\\
                  & \emph{M} -- no source.\\
                  & \emph{X} -- other issue.\\
\textbf{Severity} & \emph{Wa} -- defective data, however no repair was attempted because defect was deemed minor, i.e., did not affect the extraction.\\
                  & \emph{Mc} -- reduction procedure was modified and successfully corrected issue.\\
                  & \emph{Mi} -- reduction procedure was modified and improved the issue but could not fully correct it.\\
                  & \emph{Fa} -- there was an issue with the data which could not be improved upon.\\
\textbf{Solution} & \emph{\#Da\#} -- aperture difference (rather than nod difference). Where numbers are included in the code, the first number is the order number, and the second the nod, e.g., 1Da2 is LL1, nod 2 \\
                  & \emph{\#Dn\#} -- nod difference (rather than aperture difference).\\
                  & \emph{\#Dx\#} -- cross difference.\\
                  & \emph{\#C\#} -- co-add was changed (usually by using only one nod).
\end{tabular}
\end{center}

\cleardoublepage
\addcontentsline{toc}{chapter}{\numberline{}References}

\bibliographystyle{apj}

\end{document}